\documentclass[twoside,12pt]{article}
\usepackage{amssymb}
\newcommand{\nc}{\newcommand} 
\nc{\la}{\lambda} \nc{\La}{\Lambda} 
\nc{\al}{\alpha}
\nc{\te}{\theta}  \nc{\be}{\beta}
\nc{\ga}{\gamma}  \nc{\Ga}{\Gamma}
\nc{\de}{\delta}  \nc{\De}{\Delta}
\nc{\si}{\sigma}  \nc{\ka}{\kappa}
\nc{\om}{\omega}  \nc{\Om}{\Omega}
\nc{\nf}{\infty}   \nc{\nl}{\newline}
\nc{\ra}{\longrightarrow}
\nc{\beq}{\begin{equation}}
\nc{\eeq}{\end{equation}}
\nc{\beqa}{\begin{eqnarray}} 
\nc{\dst}{\displaystyle}	\nc{\qq}{\quad\quad}
\nc{\eeqa}{\end{eqnarray}} \nc{\nnb}{\nonumber}
\title{\bf Cohomogeneity-one Einstein\,-Weyl structures :\\ a local
approach }
\author{Guy Bonneau\thanks {\noindent Laboratoire de Physique Th\'eorique
et des Hautes Energies,
 Unit\'e associ\'ee au CNRS UMR 7589, Universit\'e Paris 7,
 2 Place Jussieu, 75251 Paris Cedex 05. bonneau@lpthe.jussieu.fr}}
\begin{document}
\date{}
\maketitle
\begin{abstract}
\noindent We analyse in a systematic way the (non-)compact n-dimensional
Einstein-Weyl spaces equipped with a cohomogeneity-one metric. In that
context, with no compactness hypothesis for the manifold on which lives
the Einstein-Weyl structure, we prove that, as soon as the
(n-1)-dimensional space is a homogeneous reductive Riemannian space
with a unimodular group of left-acting isometries G :
\begin{itemize}\item there exists a Gauduchon gauge such that the Weyl form
is co-closed and its dual is a Killing vector for the metric,
\item in that gauge, a simple constraint on the conformal scalar curvature
holds, \item a non-exact Einstein-Weyl stucture may exist only if the
(n-1)-dimensional homogeneous space G/H contains a non trivial
subgroup H' that commutes with the isotropy subgroup H,
\item the extra isometry due to this Killing vector corresponds to the
right-action of one of the generators of the algebra of the  subgroup H'.
\end{itemize} The first two results are well known when the Einstein-Weyl
structure lives on a compact manifold, but our analysis gives the first
hints on the enlargement of the symmetry due to the Einstein-Weyl
constraint.

We also prove that the subclass with G compact, a one-dimensional subgroup H' and the (n-2)-dimensional space G/(H$\times$H') being an arbitrary compact symmetric
K\"ahler coset space,
corresponds to n-dimensional Riemannian locally conformally K\"ahler
metrics. The explicit family of structures of
cohomogeneity-one under SU(n/2) being, thanks to our results, invariant
under U(1)$\times$SU(n/2), it coincides with the one first studied by Madsen~; our
analysis allows us to prove most of his conjectures. 
\end{abstract}

\vfill {\bf PAR/LPTHE/99-36/gr-qc/9912067}\hfill  November 1999
\newpage

\section{Introduction} For the last thirty years, gauge invariance has
been the guiding idea in the construction of an unified theory of all
interactions. In the genesis of the ``gauge principle", the name of the
mathematician Hermann Weyl should be recognised by physicists [the book of
L. O' Raifeartaigh \cite{OR} offers a very instructive historical review
of that subject]. The same mathematician has also defined ``Weyl geometry"
which emphasizes the role of conformal invariance : it describes not a
given metric $g$ in the target space together with a gauge field
$\ga_{\mu}$ (or a one form $\ga =
\ga_{\mu}dx^{\mu}\,)$, but an equivalence class $[g]$, through a conformal
transformation of the distance $g \rightarrow e^f g$ and a related gauge
transformation of the gauge field $\ga \rightarrow \ga + df$. So, even if
H. Weyl's original hope 
\cite{{HW29},{OR}} for an unified theory of electromagnetism and gravity
failed, it is useful to pursue some analyses of his geometry :  see some
recent efforts in the same spirit in \cite{{JTH},{WD}}.  On the other
hand, Einstein manifolds enter the game with Einstein gravity and also,
since 1969 \cite{MHE}, in the framework of the quantisation of non-linear
$\si$ models : indeed, they offer multiplicatively renormalisable 2-D
theories. Note that special Einstein manifolds are the Ricci-flat ones,
for example the  Calabi-Yau manifolds, the building block of string
theory. It is then natural to export such Einstein constraints on a Weyl
space, {\sl i. e.} to study Einstein-Weyl geometry (For a recent review see
\cite{Calderbank} and references therein).

Then, Einstein-Weyl geometry - in particular in 3 and 4 dimensions - has
raised some interest in the last years, mainly among mathematicians, but
also for physicists when 3-dimensional Einstein-Weyl geometries were used
to construct 4-D non-linear $\si$ models with (4,0) or (4,4) supersymmetry
\cite{papado-valent}, or when  Tod
\cite{Tod96} exhibited the relationship between a particular Einstein-Weyl
geometry without torsion (the four-dimensional self-dual Einstein-Weyl
 geometry studied by Pedersen and Swann
\cite{PS93}) and local heterotic geometry (i.e.  the Riemannian geometry
with torsion and three complex structures, associated with  (4,0)
supersymmetric non-linear $\si$ models
 \cite{{HullWitten85},{xxy},{Delduc}}).

In \cite{{Bonneau97},{Bonneau98}} we analysed in a systematic way, first
from a local point of view, then with completeness and compactness
restrictions, the 4-dimensional Einstein-Weyl structures equipped with a
Bianchi  metric.  This allows us to illustrate the general results
obtained by
 mathematicians around Gauduchon, Tod, Pedersen, Poon and  Swann
\cite{{Gauduchon},{Gauduchonbis},{Tod92},{PedTod92},{PDSRheine},
{PS93},{Madsen-a},{mpps}} and for example to show that Einstein-Weyl 
structures equipped with a Bianchi  metric are either conformally Einstein
or conformally K\"ahler
\cite{Bonneau97}. The aim of the present work is two-fold :
\begin{itemize}
\item A) extend our 4-dimensional study to n dimensions, still in a local
approach  and, in the spirit of 4-D separation of ``time" and ``space", we
restrict ourselves to cohomogenity-one  manifolds. In particular, we show
that the main results proved by mathematicians for {\bf compact
Einstein-Weyl structures }hold true for (non-)compact cohomogeneity-one
structures as soon as the (n-1)-dimensional principal orbit is a homogeneous
reductive Riemannian space with a unimodular group of isometries :
\begin{itemize}
\item existence of a Gauduchon gauge such that the Weyl form $\ga $ is
co-closed
\cite{Gauduchon}, and such that the dual of the Weyl form is a Killing
vector for the metric 
\cite{Tod92}, \item still in that gauge, nice constraint on the  conformal
scalar curvature
\cite{{PedTod92},{PDSRheine},{Gauduchonbis},{CalderbankFaraday}}:
$$S^D = -\frac{n(n-4)}{4}(\ga_{\nu}\ga^{\nu}) + {\rm constant}\,;$$
\end{itemize}
\item B) get a better understanding of the symmetry that corresponds to
the upper mentioned Killing vector : \begin{itemize}
\item one of our main results is a no-go theorem. If the (n-1)-dimensional
Riemannian homogeneous space is the right coset space G/H, {\bf a non exact
($\ga
\neq df$) Einstein-Weyl space exists only if there exists} a non empty
subgroup H' of G such that $H
\cap H' = \emptyset \,,\quad [H,\,H'] = 0\,.$
\item we also prove that this isometry corresponds to the right action of
one of the generators of the subgroup H', and so that {\bf the symmetry of
the solution is bigger that of the Einstein-Weyl equations :} it
is enlarged from G acting on the left ($G^L$) to $G^L
\times GL(1,\mathbb{R})\,.$ This unusual phenomenon, a kind of spontaneous
generation of symmetry, results from the Einstein-Weyl constraints : note
that such a phenomena will be helpful in the quantisation of the theory,
as a Ward identity is more manageable than a geometrical constraint such
as the Einstein-Weyl property.
\end{itemize}
\end{itemize}
 The paper is organised as follows :  in the next Section, we first 
recall the geometrical setting of Einstein-Weyl geometry and
cohomogeneity-one metrics ; then we emphasize some properties of left and
right group action on coset spaces and finally we give the
expressions of geometrical quantities, separating the n-dimensional metric
g into a ``time part" and a (n-1)-dimensional ``space part". In Section
3, focussing on unimodular groups G, we exhibit a specific Gauduchon gauge
and express the Einstein-Weyl equations in that gauge. In full generality,
we are then able to prove the announced results ( Lemma 1 and Theorem 1).
We end the Section by a characterisation of some special families of
solutions where only two functions are involved in the expression of the
Einstein-Weyl structure. In particular, we prove that for the whole family
built on an (n-2)-dimensional compact symmetric K\"ahler space, the
corresponding n-dimensional metric is locally conformally K\"ahler
(Theorem 2).  

Section 4 is then devoted to the family of  SU(m) left-invariant
structures in n = 2m dimensions. Thanks to the results of the previous
Section, the isometry group is enlarged to
$U(1) \times SU(m)^L\,.$ As in the 4-dimensional case, they are
conformally K\"ahler and we obtain the explicit expression of the
structure : it depends on 3 arbitrary parameters, up to a homothety. As
in \cite{Bonneau98}, we use the terminology of Gibbons and Hawking on nuts
and bolts \cite{GH} to search for n-dimensional regular and complete
solutions and show that, up to an arbitrary homothetic factor
$\Ga\,,$ (m + 2) one-parameter families of solutions exist. In particular,
we prove that a bolt(p)-bolt(p) solution exists iff. the twist p is 1,
2,...(m-1). This proves one of Madsen's conjectures\cite{Madsen-a}.

The same work is done in Section 5 for $S^1 \times SO(n-1)$ left-invariant
structures. We obtain the explicit expression of the structure depending
on 3 arbitrary parameters, up to a homothety. Here again, we look for
n-dimensional regular and complete solutions and show that, up to an
arbitrary homothetic factor $\Ga\,,$ only 3 one-parameter families of
non-conformally Einstein solutions exist, all with an everywhere positive
conformal scalar curvature.

Some concluding remarks are offered in Section 6. Appendix A describes the
splitting of n-dimensional geometric quantities to (n-1)-dimensional ones
for cohomogeneity-one metrics and, in appendix B, we relate two of our
families of solutions of opposite orientations.

\section{Einstein-Weyl structures \nl and cohomogeneity-one metrics :\nl
The geometrical setting}
\subsection{Weyl space} A Weyl space \cite{{Calderbank},{PS93}} is  a
conformal manifold with a torsion-free connection D and a one-form
$\ga$ such that for each representative metric g in a conformal class [g],
\beq\label{a1}
 D_{\mu} g_{\nu \rho} = \ga_{\mu} g_{\nu \rho}\ .
\eeq  \noindent A different choice of representative metric : $g\ \ra\
\tilde{g} = e^f g$ is accompanied by a change in
$\ga\ : \ga\
\ra\ \tilde{\ga} = \ga + df\ .$  Conversely, if the one-form
$\ga$ is exact, the metric g is conformally equivalent to a Riemannian
metric
$\tilde{g}$ : $ D_{\mu}\tilde{ g}_{\nu \rho} = 0.$ In that case, we shall
speak of an {\it exact} Weyl structure.

\noindent The Ricci tensor associated to the Weyl connection D is defined
by :
\beq\label{a2}  [D_{\mu},\,D_{\nu}\,]v^{\rho} = {\cal R}^{(D)\rho}_{\ \
\ \
\la,\mu\nu}\ v^{\la}\ \ ,\ \ {\cal R}^{(D)}_{\mu\nu} = {\cal
R}^{(D)\rho}_{\
\
\ \ \mu,\rho \nu}\ .
\eeq
\noindent ${\cal R}^{(D)}_{\mu\nu}$ is related to 
$R^{(\nabla)}_{\mu\nu}$, the Ricci tensor associated to the Levi-Civita
connection :
\beq\label{a3} {\cal R}^{(D)}_{\mu\nu} =  R^{(\nabla)}_{\mu\nu} +
\frac{n-1}{2}\nabla_{\nu}\ga_{\mu} -
\frac{1}{2}\nabla_{\mu}\ga_{\nu} +
\frac{n-2}{4}\ga_{\mu}\ga_{\nu} + \frac{1}{2} g_{\mu\nu}
[\nabla_{\rho}\ga^{\rho} - \frac{n-2}{2}\ga_{\rho}\ga^{\rho} ]\ .
\eeq

\noindent Using (\ref{a2},\ref{a3}), a nice relation
\cite{CalderbankFaraday} constrains the conformally invariant two-form
$d\ga$  which we call the field strength  \footnote{\, In the original
point of view of H. Weyl,
$\ga_{\mu}$ is the electromagnetic field and in
\cite{CalderbankFaraday} the term Faraday's two-form is used for  
$d\ga\,.$}
 $$d\ga =
\frac{1}{2} F_{\mu\nu}dx^{\mu}\wedge dx^{\nu}\,,$$ 
\beq\label{01} g^{\mu\la}g^{\nu\rho}D_{\la}D_{\rho}F_{\mu\nu} = 
-\frac{n-4}{4}F^{\mu\nu}F_{\mu\nu}\ \Leftrightarrow \ 
D_{\mu}D_{\nu}F^{\mu\nu} =  -\frac{n}{4}F^{\mu\nu}F_{\mu\nu}\,.
\eeq
 
\subsection{The Gauduchon gauge} In the compact case, up to a homothety
there exists a unique metric
$g$ in the conformal class  such that $\ga$ is co-closed :
$$\nabla_{\la}\ga^{\la} = 0\,.$$ (The Lorentz gauge for electromagnetism.)

\subsection{Einstein-Weyl spaces} Einstein-Weyl spaces are Weyl structures
defined by 
\footnote{\, [a,b] and (a,b) respectively mean antisymmetrisation  and
symmetrisation with respect to the indices a,b : $v_{(a} w_{b)} = [v_a w_b
+v_b w_a]/2$, e.t.c...} :
\beqa\label{a5}  {\cal R}^{(D)}_{(\mu\nu)}  & = &
\frac{S^D}{n}\ g_{\mu\nu}\
\ \Leftrightarrow \nnb \\
 R^{(\nabla)}_{\mu\nu} +  \frac{n-2}{2}[\nabla_{(\mu}\ga_{\nu)} +
\frac{1}{2}\ga_{\mu}\ga_{\nu}] & = & \La\  g_{\mu\nu}\ \ ,\ \
\La  =
\frac{S^D}{n} - \frac{1}{2}[\nabla_{\la}\ga^{\la} -
\frac{n-2}{2}\ga_{\la}\ga^{\la}] \ .
\eeqa 
\noindent Note that for an exact Einstein-Weyl structure, $\ga = df$, the
representative metric is conformally Einstein. Note also that the
conformal scalar curvature is related to the scalar curvature through:
\beq\label{a6}  S^{D} = g^{\mu\nu}{\cal R}^{(D)}_{\mu\nu}  = n\La +
\frac{n}{2}[\nabla_{\la}\ga^{\la} -
\frac{n-2}{2}\ga_{\la}\ga^{\la}] = R^{(\nabla)} +
(n-1)[\nabla_{\la}\ga^{\la} -
\frac{n-2}{4}\ga_{\la}\ga^{\la}]\ .
\eeq

\noindent For any Einstein-Weyl structure, another nice relation may be
derived using the Bianchi identity:
\beq\label{02} - \nabla_{\nu}\left[\frac{S^D}{n} +
\frac{n-4}{4}\ga_{\la}\ga^{\la}\right]  + (\nabla_{\nu}
-\frac{1}{2}\ga_{\nu})(\nabla_{\la}\ga^{\la}) =  (\nabla^{\la}
+\ga^{\la})(\nabla_{(\la}\ga_{\nu)})\,.\eeq  

\noindent Notice that in a Gauduchon gauge and when the manifold is
compact \footnote{\ At least, as soon as integration by parts on the
manifold is possible.}, contraction of (\ref{02}) with
$\ga^{\nu}$, followed by an integration on the manifold, ensures that the
vector
$\ga_{\la}$, dual of the Weyl form $\ga$, is a Killing vector
\cite{Tod92}.

\noindent A related relation is 
\cite{{PedTod92},{Gauduchonbis},{CalderbankFaraday}} :
\beq\label{02b} \left( \nabla_{\nu} + \ga_{\nu}  \right)\frac{S^D}{n} +
\frac{1}{2}\,g^{\la\mu}D_{\la} F_{\mu\nu} = 0\,.\eeq

\subsection {Cohomogeneity-one metrics}  Cohomogeneity-one metrics are
real n-dimensional metrics with an isometry group G whose generic orbits are
(n-1)-surfaces (we also restrict ourselves to  effectively acting
isometries, {\sl i.e.} the isotropy subgroup H contains no non-trivial normal subgroups, discrete or not, of G \cite{Besse}).
This  generalises to n-dimensions  the homogeneity property of
3-dimensional ordinary space in  gravity and the n-dimensional distance is
then split as :
\beq\label{2-11} ds^2  = g_{\mu
\nu}dx^{\mu}dx^{\nu} = (dT)^2 + (d\tau)^2 = (dT)^2 +  
g_{\al\be}dx^{\al}dx^{\be}\
\ \ \mu,\ \nu = (0,\al),\ (0,\be)\,,
\eeq 
 \noindent where, given some  ``proper time " $T,$ the
$T$-fixed (n-1)-space will be a homogeneous space, {\it i.e.} a
coset space G/H with G a connected group and  H a closed subgroup. As we
consider only Riemannian manifolds, the isotropy subgroup H, being a
subgroup of some orthogonal group, is compact \cite{Friedan}.  The
compactness of H ensures that G/H is a reductive homogeneous space
\cite{KoNo}, {\sl i. e.}, ${\cal G}$ and ${\cal H}$ denoting the
corresponding Lie algebras, an invariant, non-degenerate, bilinear
quadratic form on ${\cal G}$ exists and ${\cal G}$ may be decomposed
according to 
 $${\cal G = H} \oplus {\cal M} $$ where ${\cal M}$ is Ad(H)  invariant.
So the commutation relations write  :
\beqa\label{010}
\left[ h_a,\,h_b\right]& = & f_{ab}^c h_c\,,\ \ \ h_a \in {\cal H}\;;\ \
\ a,\,b,\,c = 1\,,2\,,..,L,\ \ {\rm dim}\ H = L\,, \nnb\\
\left[ h_a,\,W_i\right]& = & f_{ai}^j W_j\,,\ \ \ W_i \in {\cal M}\;;\ \
\ i,\,j,\,k = 1\,,2\,,..,(n-1), \\
\left[ W_i,\,W_j\right]& = & f_{ij}^k W_j + f_{ij}^c h_c \,.\nnb \eeqa

\noindent A parametrisation of the (n-1)-homogeneous space  is
conveniently done through (right) equivalence classes in G/H in one to one
correspondance with the considered point $x$ on the (n-1)-surface :

$$ \left[L(x)\right] \in G\,;\ \ L(x) \sim L'(x)  \Leftrightarrow \
\exists h \in H,\ / \ L(x)= L'(x).h$$  
\noindent The left action of an arbitrary $g_0$ writes :
$$\left[L(x')\right] = \left[g_0 .L(x)\right] \Leftrightarrow L(x') = g_0
.L(x).h^{-1}[x,\,g_0]\,;$$

\noindent  note that a left $h_0$ transformation, given by $L(x') =
h_0.L(x).h_0^{-1}\,,$ acts linearly on
$x\,.$

The Lie-algebra valued Maurer-Cartan one-form 
$$ M = L^{-1}(x)dL(x) $$ \noindent defines the one-forms
$e^i(x)$ and
$\om^a(x)$  by :
\beq\label{ew0} M =  e^i(x) W_i  + \om^a(x)  h_a\,;\eeq
\noindent  in particular, the one-forms $e^i(x) = e^i_{\al}(x)dx^{\al}$
transform, under an  arbitrary transformation of G, in an ``homogeneous"
way, according to $$ e^i(x)
 W_i \ra  e^i(x') W_i = h[x,\,g_0]. e^i(x)W_i .h^{-1}[x,\,g_0]\,.$$ The
infinitesimal version writes :\beq\label{0100}
\de_{g_0} e^i(x) = - \epsilon^a(x,g_0)f_{aj}^i e^j\,,\ \ {\rm with\ }
h[x,\,g_0] = \exp[- \epsilon^a(x,g_0)h_a]\ {\rm and}\ \epsilon^a(x,h_0)
\equiv \epsilon^a(h_0)\,.
\eeq 
\noindent Then, the most general G-invariant distance on the
(n-1)-dimensional space may be written as\,:\beq\label{011} (d\tau)^2 =
h_{ij}e^i(x)e^j(x)
\equiv  h_{ij}e^i_{\al}(x)e^j_{\be}(x)dx^{\al}dx^{\be}\,, \eeq 
 where the symmetric, positive definite $(n-1)\times (n-1)$ tensor
$h_{ij}$ has to be invariant under H, {\it i.e. :}
\beq\label{012}
 f_{ai}^k h_{kj} + f_{aj}^k h_{ik} = 0\,,
\eeq 
\noindent and the $e^i_{\al}(x)$ are some vielbeins. Let $\eta_{ij}$ be
independent solutions  of (\ref{012}) : they correspond to irreducible
orthogonal representations  (irreps) of the compact group H and may be used
to write the H-invariant 2-tensor
$h_{ij}$ in block-diagonal form :
 \beq\label{2-110b} d\tau^2  =  \sum_{\eta\ = irreps\ of\ H}
h^{\eta}\eta_{ij}e^i(x) e^j(x)\
\, \eeq
\noindent where $\eta_{ij}$ is a positive definite symmetric  matrix in
the irreducible component labelled by $\eta\,,$ and the $h^{\eta}$\,'s are
some arbitrary positive parameters. The cohomogeneity-one requirement
means that at any ``proper time" $T\,,$ the (n-1)-dimensional distance
takes the form (\ref{011},\ref{2-110b}) :
$$(ds)^2 = (dT)^2 + (d\tau)^2 = (dT)^2 + h_{ij}[T]e^i(x)e^j(x) = (dT)^2 +
\sum_{\eta\ = irreps\ of\ H} h^{\eta}[T]\eta_{ij}e^i(x) e^j(x)\,.$$
\noindent   Notice that there is no loss of generality in choosing the
metric element
$g_{00} = 1$ as this corresponds to a choice of "proper time " T. The
general analysis of Einstein-Weyl equations will use G-invariant
cohomogeneity-one Weyl structure written as :
\beq\label{2-110} ds^2  =  dT^2 + h_{ij}(T)e^i e^j\
\  ;  \ \ \ga  =  \ga_0(T)dT + \ga_i(T) e^i\,,\eeq 
\noindent with ($h^{ij}$  is the matrix inverse of
$h_{ij}$) :
\beqa\label{2-110a} a) & f_{ai}^k h_{kj}[T] + f_{aj}^k h_{ik}[T] & = 
0\,,\nnb
\\ b) & f_{ai}^j
\ga_j(T) = 0 & \stackrel{(\ref{2-110a}a)}{\Leftrightarrow} \ \ f_{ai}^j
\ga^i(T) = 0\,, \
\ga^i = h^{ij}\ga_j\,.\eeqa

\noindent Some inverse vielbeins 
$E_i^{\al}$ may be defined by :
\beq\label{2-111} dx^{\al} = E^{\al}_i e^i\ \Rightarrow \
e^i_{\al}E_j^{\al} =
\de_{j}^{i}
\,,
\ e^i_{\al}E_i^{\be} =
\de_{\al}^{\be}\,.
\eeq 
\noindent The Maurer-Cartan consistency condition $ dM + M \wedge M = 0$
gives :

\beqa\label{014}  de^i +\frac{1}{2} f_{jk}^{i}e^j\,\wedge e^k +
f_{ak}^{i}\om^a\,\wedge e^k = 0 & \Rightarrow & 
\nabla_{[\be}e^i_{\al]} =
\frac{1}{2}f_{jk}^i e^j_{\al} e^k_{\be} +  f_{a\,k}^i \om^a_{[\al}
e^k_{\be]} \,,\\ d\om^a +\frac{1}{2} f_{jk}^{a}e^j\,\wedge e^k +
\frac{1}{2}f_{bc}^{a}\om^b\,\wedge \om^c = 0 & \Rightarrow  &
\nabla_{[\be}\om^a_{\al]} =
\frac{1}{2}f_{jk}^a e^j_{\al} e^k_{\be} +  \frac{1}{2} f_{b\,c}^a
\om^b_{\al}
\om^c_{\be}\,.\nnb
\eeqa 
\noindent Moreover (see Appendix A), using equations
(\ref{012},\ref{014}), one obtains for the symmetrised  covariant
derivative  :
\beq\label{015}
\nabla_{(\be}e^i_{\al)} = - h^{ij}f_{j(l}^kh_{m)k} e^l_{\al} e^m_{\be} -
f^i_{am}\om^a_{(\al}e^m_{\be)}\,.
\eeq

\subsection{The right action of G} For further use, note that a right
action of G on right equivalent classes in G/H can  be defined for
those elements h' of G (with h' not in H) that  commute with all elements
of H \footnote{\ Of course, a right action of all elements of G that normalise H may always be defined. However, in the analysis of the isometries of Einstein-Weyl structures, only those elements that commute with H will play a role as the corresponding one-forms are left-invariant.} . They form a  subgroup H' of G. The corresponding Lie-algebra
elements belong to $({\cal G - H})$ and generate a sub-algebra
${\cal H'}$ ; the Lie-algebra of G may be decomposed according to
$${\cal G = H} \oplus {\cal H'} \oplus {\cal M} $$ 
\noindent where ${\cal M}$ is the complement of
${\cal H} \oplus {\cal H'}$.  The commutation relations are now :
\beqa\label{010b}
\left[ h_a,\,h_b\right]& = & f_{ab}^c h_c\,,\ \ \ h_a \in {\cal H}\;;\ \
\ a,\,b,\,c = 1\,,2\,,..,L,\ \ {\rm dim}\ H = L\,, \nnb \\
\left[ h'_u,\,h'_v\right]& = & f_{uv}^w h'_w\,,\ \ \ h'_u \in {\cal
H'}\;;\ \
\ u,\,v,\,w = 1\,,2\,,..,L',\ \ {\rm dim}\ H' = L'\,, \nnb \\
\left[h_a,\,h'_u \right] & = & 0, \nnb \\
\left[ h_a,\,\tilde{W}_i\right]& = & f_{ai}^j \tilde{W}_j\,,\ \ \
\tilde{W}_i
\in {\cal M}\;;\ \
\ i,\,j,\,k = 1\,,2\,,..,(n-1 - L'), \\
\left[ h'_u,\,\tilde{W}_i\right]& = & f_{ui}^j \tilde{W}_j\,\nnb \\
\left[ \tilde{W}_i,\,\tilde{W}_j\right]& = & f_{ij}^k \tilde{W}_j +
f_{ij}^c h_c + f_{ij}^uh'_u\,.\nnb
\eeqa  
\noindent In that case,  note that this right action of $h'$ on right
equivalence classes is simply defined by : 
$$ \left[\tilde{L}(x')\right] = \left[\tilde{L}(x).h'_0\right] =
\left[\tilde{L}(x)\right].h'_0\,. $$ Moreover, the Maurer-Cartan one-form
M should now be decomposed as 
\beq\label{010c}  M =  \tilde{e}^i(x)\,\tilde{W}_i +  y^u(x)\,h'_u + 
\tilde{\om}^a(x)\,h_a
\,.\eeq 
\noindent On the one hand, note that the one-forms $y^u(x)$ are
left-invariant and the G-invariant  cohomogeneity-one Weyl structure may be
written as \footnote{\ Using irreducible representations of H (see the
previous subsection), $\tilde{h}_{ij}$ could be written as in
(\ref{2-110b}) ; moreover 
$\tilde{h}_{uv}$ is an arbitrary positive definite symmetric matrix.} :
\beq\label{2-113} ds^2  =  dT^2 + \tilde{h}_{ij}(T)\tilde{e}^i\tilde{e}^j + 
\tilde{h}_{uv}(T) y^u y^v\
\  ;  \ \ \ga  =  \ga_0(T)dT + \tilde{\ga}_u(T) y^u \,,\eeq 
\beq\label{2-113a}  {\rm with}\,:\ f_{ai}^k \tilde{h}_{kj}[T] + f_{aj}^k
\tilde{h}_{ik}[T]
 =  0\,.\eeq

\noindent On the other hand, under a right action, the one-forms
$\tilde{e}^i(x)$ and $ y^u(x)$ transform linearly, and the $\om^a (x) $
are invariant :
$$\delta_{h'}\tilde{e}^{i}(x) = - \epsilon^{u} f_{u\,j}^{i}
\tilde{e}^{j}(x)\,,
\ \ \delta_{h'}y^{v}(x) = - \epsilon^{u} f_{u\,w}^{v}\,y^w(x)\,;$$  
\noindent then, the Weyl structure (\ref{2-113},\ref{2-113a}) remains of
the same form ( with a tensor
$\tilde{h}_{ij}[T]$ changed into another H
invariant one, thanks to Jacobi identities) according to :
\beqa\label{x1} \tilde{h}'_{ij} & = & \tilde{h}_{ij} +
\epsilon^{u}[f_{u\,i}^{k}\tilde{h}_{kj} + f_{u\,j}^{k}\tilde{h}_{ik}] \
\,,\
\tilde{h}'_{uv} = \tilde{h}_{uv} +
\epsilon^{w}[f_{w\,u}^{u'}\tilde{h}_{u'\,v} +
f_{w\,v}^{u'}\tilde{h}_{u\,u'}]
\,,\\ 
\tilde{\ga}'_u & = & \tilde{\ga}_u  +
\epsilon^{v}f_{v\,u}^{w}\tilde{\ga}_w \,.\nnb\eeqa

\noindent (A discussion of non-linear $\sigma$ models built on homogeneous spaces with such a H' subgroup may
be found in subsections 3.1-2 of
\cite{BBBCD}) .

\subsection{n-dimensional geometric quantities} The n-dimensional
geometric quantities  may now be expressed as functions of the
(n-1)-dimensional ones (Appendix A).

First, thanks to previous results (\ref{014}-\ref{015}),
$R^{(\nabla)}_{\mu\nu}$ may be expressed as (for example see
\cite{{Landau},{Bonneau97}})~:
\beqa\label{2-12} R^{(\nabla)}_{00} & = &
-\frac{1}{2}\frac{d}{dT}(\frac{h'}{h}) -
\frac{1}{4}K_i^j K_j^i\ \ ;\ K_i^j =
\frac{dh_{ik}}{dT}h^{kj}\ \ ;\ h =
\det[h_{ij}]\ , h' = \frac{dh}{dT}\ , \nnb\\ R^{(\nabla)}_{0\al} & = &
\frac{1}{2} e_{\al}^k [f^i_{kj} - \de^i_k f^m_{jm}] K_i^j\ ,\\
R^{(\nabla)}_{\al\be} & = &
\si_{\al}^i\si_{\be}^j\left[ R^{(n-1)}_{ij} -
\frac{1}{2}\frac{dK_{ij}}{dT} + \frac{1}{2}K_i^k K_{kj} -
\frac{h'}{4h}K_{ij}\right]\ ;\ K_{ij} = K_i^k h_{kj} =
\frac{dh_{ij}}{dT}\ ,\ {\rm e.t.c...}\nnb
\eeqa 
\noindent where $ R^{(n-1)}_{ij}$, the (n-1)-dimensional Ricci tensor
associated to the homogeneous space Levi-Civita connection, in the
vielbein basis
$e^i$, may be expressed as a function of the metric
$h_{ij}$ and of the structure constants of the group
\cite{{Landau},{russes}}.

 Second, the Bianchi identity splits \cite{Bonneau97} :
\beq\label{2-13a} f^i_{jk}R^{(n-1)j}_i + f^i_{ji}R^{(n-1)j}_k = 0\ \ ,\ k
=1,2,..,n-1\ \
\ \cite[\rm equ.(116,25)]{Landau}\,,
\eeq and :
\beqa\label{2-13b} h^{ij}\frac{d}{dT}R^{(n-1)}_{ij}  \equiv 
\frac{dR^{(n-1)}}{dT} + K_i^jR^{(n-1)i}_j & = & 2
(\nabla_{\al}E^{\al}_i)R_0^i\ ,\\  {\rm with}\ :\ \nabla_{\al}E^{\al}_i =
f_{ji}^i + (\om^a_{\be}E^{\be}_l)\,f_{a\,i}^l\,,\ \ R_0^i & = &
h^{ij}E^{\al}_j R^{(\nabla)}_{0\al}\
\ ,\ R^{(n-1)} = R^{(n-1)}_{ij}h^{ij}\,.\nnb
\eeqa \noindent We do not find the nice equation (\ref{2-13b}) in the
standard textbooks on gravity (even for ordinary 4-dimensional space time
with a 3-d group of isometries, $i.e.$ no subgroup H, where
$\nabla_{\al}E^{\al}_i$ simplifies to $ f_{ji}^i\,.$ ).

Third, one may also obtain using (\ref{2-110a},\ref{014},\ref{015}):
\beqa\label{2-13c}
\nabla_0\ga_0 & = & \frac{d\ga_0}{dT}\ ,\nnb \\
\nabla_{(0}\ga_{\al)} & = & \frac{1}{2}e_{\al}^i h_{ij}\frac{d\ga^j}{dT}\
,\
\ga^i = h^{ij}\ga_j\ ,\nnb \\
\nabla_{(\al}\ga_{\be)} & = & e_{\al}^i e_{\be}^j[\frac{1}{2}\ga_0 K_{ij} +
 h_{l(i}f^l_{j)k}\ga^k]\ ,\\ \nabla_{\mu}\ga^{\mu} & = &
\frac{d\ga_0}{dT} +\frac{h'}{2h}\ga_0 +\ga^i f_{ji}^j \ \ .\nnb\eeqa
\noindent  Note that one may always choose a representative in the
conformal class [g] such that
$\ga_0 (T)
\equiv 0\,.$ Then, as soon as $f_{ij}^j = 0\,,i = 1,\,2,..(n-1),$
\footnote{ As $f_{i\,a}^{a} =0\,$ and ( thanks to the compactness of H)
$f_{a\,j}^{j}+ f_{a\,b}^b = 0\,,$ our condition reduces itself to the
unimodularity of the adjoint action of G. This also means that the 
measure is G-invariant.} , this choice gives a special family of Gauduchon
gauges
\cite{Bonneau97}, and, in the rest of this study, we suppose that this
condition is fulfilled.

\section{The Einstein-Weyl equations \nl in the special gauge
$\ga_0 = 0$}
\subsection{General results} For cohomogeneity-one metrics, the
Einstein-Weyl equations (\ref{a5})  may be split and written in the
special gauge
$\ga_0 = 0$ ( let us recall that we consider algebras ${\cal G}$ with
$f_{ij}^j = 0$) :
\beq\label{ew1}
\La  =  -\frac{1}{2}\frac{d}{dT}(\frac{h'}{h}) -
\frac{1}{4}K_i^j K_j^i\  ;\eeq
\beq\label{ew2} 0  = 
\frac{1}{2}f^i_{kj} K_i^j +\frac{n-2}{4}h_{ki}\frac{d\ga^i}{dT}\ ;\eeq
\beq\label{ew3} h_{ij}\La  = 
 R^{(n-1)}_{ij} -
\frac{1}{2}\frac{dK_{ij}}{dT} + \frac{1}{2}K_i^k K_{kj} -
\frac{h'}{4h}K_{ij} + \frac{n-2}{2}\ga^k h_{l(i}f^l_{j)k} +
\frac{n-2}{4}\ga_i\ga_j\ .\eeq

\noindent On the one hand, the use of relations
(\ref{ew1},\ref{ew2},\ref{2-13b}) in the equations obtained through
contraction of (\ref{ew3}) with
$h^{ij}$ and
$K^{ij}\,,$  gives :
\beq\label{ew4}
\frac{d}{dT}\left[S^D +Ê\frac{n(n-4)}{4}\ga_i\ga^i\right] =
-\frac{n}{2}\frac{d\ga^i}{dT}\left[\nabla_{\al}E^{\al}_i -
\frac{(n-4)}{2}\ga_i \right]\,.\eeq On the other hand, equation (\ref{02})
splits into :
\beq\label{ew5}
\frac{d}{dT}\left[S^D +\,\frac{n(n-4)}{4}\ga_i\ga^i\right] = 
-\frac{n}{2}\frac{d\ga^i}{dT}\left[\nabla_{\al}E^{\al}_i + \ga_i
\right]\,,\eeq and 
\beq\label{ew6}
\frac{d}{dT}[h_{ij}\frac{d\ga^j}{dT}] = \ga^j[f_{ji}^k\ga_k + X_{ij} +
\nabla_{\al}E^{\al}_k (D_j)^k_i]\,,\eeq where the traceless matrices
$D_i$ have for matrix elements
$(D_i)_m^n \equiv f_{im}^{\,n} + f_{ir}^{\,s} h_{sm}h^{rn}\,$ and
$X_{ij}$ is a symmetric, non-negative matrix :
$$X_{ij} = f_{im}^n [ f_{jn}^m + f_{jr}^s h_{ns}h^{mr}] =  2f_{i(m}^{\,n}
h_{s)n}f_{jr}^s h^{rm}\, = 1/2 (D_i)_m^n(D_j)_n^m\,.$$ Indeed, V being any
eigen-vector of the symmetric matrix $X$ with eigen-value
$\la$,
$$\sum_j X_{ij} (V)_j =  \la(V)_i\,,$$ one can choose a diagonal basis for
the positive definite metric
$h_{ij}$ and compute
$$ \la \sum_i(V)_i(V)_i = \sum_{i,j}X_{ij}(V)_i(V)_j = 1/2
\sum_{s,t}h^{ss}h_{tt} [\sum_i(V)_i (D_i)]_s^t [\sum_j(V)_j (D_j)]_s^t
\ge 0\,.$$ As a consequence, the eigen-values of the matrix $X$ are
non-negative.\hfill Q.E.D.\nl\noindent Moreover, the existence of a zero
eigen-value requires an eigen-vector $V$ satisfying :
$$[\sum_i(V)_i (D_i)]_s^t = 0 \Leftrightarrow \sum_i(V)_if_{i(m}^s h_{n)s}
= 0\,.$$

\noindent Equations (\ref{ew4},\ref{ew5}) give (as soon as $n \ge 3$) :
\beq\label{ew7} \ga_i\frac{d\ga^i}{dT} = 0\,.\eeq Then, contraction of
(\ref{ew6}) with $\ga^i$ using (\ref{2-110a}) and
$\nabla_{\al}E^{\al}_i = (\om^a_{\be}E^{\be}_l)\,f_{a\,i}^{l}$ leads to
:\beq\label{ew8}
\frac{d\ga^i}{dT}h_{ij}\frac{d\ga^j}{dT} + \ga^i X_{ij}\ga^j = 0
\,,\eeq which enforces \beq\label{ew41}
\frac{d\ga^i}{dT} = 0 \Leftrightarrow \ga^i =
\Ga^i\ {\rm constants\ constrained\ by}\ (\ref{2-110a})\ :\ \Ga^i f_{ai}^j
= 0\,. \eeq  As a consequence, the operator \beq\label{2-110c} Z_{\Ga} =
\Ga^i W_i\eeq
 commutes with all $\ h_a\,.$

\noindent Then, a non exact \footnote{\,An exact Einstein-Weyl structure
with a non-vanishing $\ga$ also requires the existence of some h' element
in the algebra $({\cal G - H})\,,$ but in that work, we consider mainly
non-exact  Einstein-Weyl structures.} Einstein-Weyl structure, {\it i.e.}
a solution with at least one non vanishing H-invariant (n-1)-vector 
$\Ga^ i $, requires that the algebra
$({\cal G - H})$ contain some h' element (see the Subsection 2.5) 
and that there exist at least one zero eigen-value for
$X\,:$

\beq\label{ew9}
\Ga^i(D_i)_m^s\,h_{ns} = \Ga^i f_{i(m}^s h_{n)s} = 0\ .\eeq 

Then we have~:\nl

\vspace{3mm}\noindent {\bf Lemma 1 :} {\it  Given a homogeneous
(n-1)-dimensional space G/H, the Lie algebra of G satisfying
$\dst\sum_j\,f_{ij}^j = 0,\ i,j\,=1,..n-1\,,$ a n-dimensional non-exact
Einstein-Weyl structure of cohomogeneity-one under the left-action of G
may exist {\bf only if} at least one generator in ${\cal G - H}$ commutes
with all the generators of
${\cal H}$.}
\vspace{3mm}

In particular, as the structure constants of a symmetric coset space
satisfy $f_{ij}^k = 0\,,$ one gets~: 

\vspace{3mm}
\noindent {\bf Corollary 1 :} {\it Any n-dimensional  Einstein-Weyl
structure of cohomogeneity-one under the action of a group G and  whose principal orbit G/H is a symmetric space without flat factors, can only be an
exact Einstein-Weyl structure.}

\vspace{3mm} As a particular case, note that $G = SO(n)$ being the maximal
isometry group of an (n-1)-dimensional homogeneous space 
 \cite{W}, in that situation G/H will be the sphere
$S^{n-1}\,;$  but, an $SO(n-1)$ invariant Weyl form $\ga$ reduces to
$\ga_0 dT\,.$ So, the sole Weyl structure is an exact one, in agreement
with Corollary 1.

\vspace{3mm} Thanks to the discussion in Subsection 2.5, a right
action of
$Z_{\Ga}$ may then be defined. Under an infinitesimal right group
transformation $Z^R =
\exp[\epsilon Z_{\Ga}]\,,$ any representative of a right equivalence class
in G/H transforms according to $$L(x') = L(x)\,.Z^R\,.h^{-1}(x,\Ga)\,,$$ 
the one-forms $e^i(x)$ and $\om^a(x)\,,$ still defined through
(\ref{ew0}),  transform  according to :
\beqa\label{ew351} \de_{Z^R}\,e^i(x) & = & -
 [\epsilon^b(x,\Ga)\,f_{b\,j}^{i} +
\epsilon\,\Ga^k\,f_{k\,j}^{i}]e^j(x)\,,\\
\de_{Z^R}\, \om^a(x) & = & -
 [\epsilon^b(x,\Ga)\,f_{b\,c}^{a}\om^c(x) +
\epsilon\,\Ga^k\,f_{k\,j}^{a}e^j(x)] + d\epsilon^a(x,\Ga)\,,\ {\rm where}\
h(x,\Ga) =
\exp[-\epsilon^a(x,\Ga)h_a]\,.\nnb \eeqa  The ``gauge" function
$\epsilon^a(x,\Ga)$ can be expressed as 
\beq\label{ew352}
\epsilon^a(x,\Ga) = \epsilon \Ga^i\,\om_{\al}^a(x)\,E_i^{\al}(x)\;;\eeq

\noindent Indeed,  
$$L^{-1}(x).\left[L(x + \delta x) - L(x)\right] =
Z^R\,.\exp[\epsilon^a(x,\Ga)h_a] - 1 \simeq \epsilon\Ga^k W_k +
\epsilon^a(x,\Ga)h_a \,,$$ and, when one uses the Maurer-Cartan one form
M(x), the left hand side expression writes :
$$L^{-1}(x).\left[L(x + \delta x) - L(x)\right] \simeq [e^i_{\al}(x) W_i +
\om^a_{\al}(x) h_a]\delta x^{\al}\,;$$ identification allows the
elimination of $\delta x^{\al} $ and gives the announced result
(\ref{ew352}).

Using (\ref{2-110a},\ref{ew9}), the distance and Weyl form are readily
checked to be invariant, which shows that the symmetry group of the
Einstein-Weyl structure is enlarged from
$G^L $ to $G^L\,\times\,GL(1,\mathbb{R})\,,$ as there exists a combination
of the left and right action of $Z_{\Ga}$ which acts linearly
\cite{BBBCD}

\vspace{5mm} 

Let us now make contact with the general results obtained for a
compact n-dimensional Einstein-Weyl structure in the (unique) Gauduchon
gauge. First, equation (\ref{ew5}) gives, in agreement with
\cite{{PDSRheine},{Gauduchonbis},{CalderbankFaraday}} :
\beq\label{ew10} S^D +\frac{n(n-4)}{4}\Ga^ih_{ij}(T)\Ga^i = {\rm
constant}\,;\eeq second, relations (\ref{2-13c}) and (\ref{ew9}) lead to
$\nabla_{(\mu}\ga_{\nu)}  = 0\,,$ and enforce $ \ga^{\mu}$ to be a Killing
vector for the metric, in agreement with
\cite{Tod92}, the corresponding isometry generator being 
\beq\label{ew82}
\tilde{Z}_{\Ga} = \Ga^i E_i^{\al}\frac{\partial}{\partial x^{\al}}\,.\eeq 
$$    $$ Note that $\nabla_{(\al}\ga_{\be)}  = 0$ also enforces
$ \ga^{\al}$ to be a Killing vector on T-fixed surfaces. Thanks to
equation (\ref{014}),
$\tilde{Z}_{\Ga}$ acts on one-forms $e^i(x)$ and
 $\om^a(x)$ according to :
\beqa\label{ew81}
\tilde{Z}_{\Ga}e^i & = & -\left[\Ga^m f_{mj}^i + (\Ga^m
E_m^{\al}\om_{\al}^b)f_{bj}^i\right]e^j\,,\nnb\\
 \tilde{Z}_{\Ga}\om^a & = & -
 \left[ (\Ga^m E_m^{\al}\om_{\al}^b)f_{b\,c}^{a}\om^c +  \Ga^m
f_{k\,j}^{a}e^j \right] + d(\Ga^m E_m^{\al}\om_{\al}^a) \eeqa and leaves
the Weyl-form invariant as :
$$\tilde{Z}_{\Ga}[\Ga^j\,h_{ij}e^i] =  \Ga^j\Ga^m\left[h_{ij}f_{mn}^i +
E_m^{\al}\om_{\al}^a\,h_{ij}\,f_{an}^i\right]e^n =
-\Ga^j\Ga^m\,h_{in}f_{mj}^i e^n = 0\,.$$ The identification
$\exp[\epsilon\tilde{Z}_{\Ga}] = Z^R$ immediatly results when one compares
equations (\ref{ew351}-\ref{ew352}) and (\ref{ew81}).

\vspace{3mm} So we have, with the notations of equations
(\ref{010},\ref{ew0},\ref{2-110})~:

\vspace{3mm}
\noindent {\bf Theorem 1 :} {\it Given a  reductive  homogeneous (n-1)
dimensional space G/H, where H is a closed subgroup of the connected group
G and G is not necessarily compact but its regular representation is
supposed to be unimodular :
\begin{itemize}
\item i) a n-dimensional non-exact Einstein-Weyl structure of
cohomogeneity-one under the left-action of G  may exist only if some
generators of $({\cal G - H})$ commute with all the generators of
${\cal H}$ (let  ${\cal H'}$ be the subalgebra of such generators,  and L'
its dimension) ;
\item ii)  $ h'_{0}\,,$ one of the generators of ${\cal H'}\,,$ being
chosen, in the particular Gauduchon gauge obtained for
$\ga_0 = 0$ the isometry group contains an extra $GL(1,\mathbb{ R})\,,$ 
corresponding to a right-action of $h'_0\,;$
\item iii) in that gauge, the Weyl form is dual to the Killing vector of
the chosen $h'_0\,;$ it is then  given by
$\ga =
\Ga^i_{0}\, h^{0}_{ij}(T) e^j\,,$ where the
$\Ga^i_{0}$ are constant parameters constrained by
$\Ga^i_{0}\,f_{ai}^j = 0\,, $ and the distance is written as 
$$ (ds)^2 = (dT)^2 + h_{ij}^0[T]e^i\,e^j \,;$$
\item iv)  the $GL(1,\mathbb{ R})\times H$-invariant metric
$h_{ij}^0[T]$ is constrained by 
$f_{a(i}^k\,h_{j)k} =  \Ga^i_0\,f_{i(m}^k\,h_{n)k} = 0$ and by the
equations :
\beqa\label{ew11} a) & \La' & =  \ -\frac{1}{2}\Ga^i
\Ga^j\,h_{ij} -\frac{1}{2}\frac{d}{dT}(\frac{h'}{h}) -
\frac{1}{4}K_i^j K_j^i\  = {\rm constant}\nnb\\ b)  & f_{ij}^k K_k^j & =
\  0\,,\\ c) & R^{(n-1)}_{ij} & = \
\La'\,h_{ij} +
\frac{1}{2}\frac{dK_{ij}}{dT} - \frac{1}{2}K_i^k K_{kj} +
\frac{h'}{4h}K_{ij} +\frac{1}{2}\Ga^m_0
\Ga^n_0[h_{mn}\,h_{ij} - \frac{n-2}{2}h_{mi}\,h_{nj}\,] \,; \nnb \eeqa
\item vi) still in that gauge, the conformal scalar curvature satisfies :
\beq\label{ew101} S^D +\frac{n(n-4)}{4}\Ga^i_{0}\,h^{0}_{ij}(T)\Ga_{0}^j =
n\La' \eeq
 \item vii)  as explained in Subsection (2.5), the (L' - 1) extra
generators of the subgroup H' offer right-transformations from one
solution with some
\{$\,h_{ij}[T]\,,\ \ga_i[T]$\} to another solution : although not
conformally equivalent, these solutions are related and should be considered as
physically equivalent.
\end{itemize}}
\noindent Let us now use the notations of equations 
(\ref{2-110b},\ref{010b}-\ref{x1}) and select
$h'_{u_0}\,,$ one of the generators of
${\cal H'}$; let
$y^{u_0}$ be the corresponding one-form vielbein defined through the
Maurer-Cartan one-form M (\ref{010c}). The $G^L\times GL(1,\mathbb{R})$
invariant Einstein Weyl structure may be rewritten - using irreducible
representations of $H \times GL(1,\mathbb{ R})$ - in a block-diagonal form
:
\beqa\label{ew110} (ds)^2 & = & (dT)^2  +
\tilde{h}^{u_0}_{ij}[T]\tilde{e}^i\tilde{e}^j +
\tilde{h}_{0}[T]y^{u_0}\,y^{u_0} + \tilde{h}_{uv}[T]y^u\,y^v\,,\ \
\ga = \Ga_{u_0}\tilde{h}_0[T]\,y^{u_0}\,,\\ u,\,v\,& = &  1,..,L'-1\,;\
\ i,\,j\,=1,...,n-1-L'\,,\nnb\eeqa
 where :\begin{itemize}
\item $\Ga_{u_0}$ is an arbitrary real parameter,
\item $\tilde{h}_{0}[T]$ is an arbitrary positive function,
\item $\tilde{h}^{u_0}_{ij}$ is a symmetric
$(n-1-L')\times(n-1-L')$ 2-tensor, invariant under
$\tilde{H} = H\times\,GL(1,\mathbb{R})\,,$
$$\tilde{h}^{u_0}_{ij}[T]e^i(x) e^j(x) =\sum_{\eta\ = irreps\ of\
\tilde{H}}
\tilde{h}^{\eta}[T]\eta_{ij}e^i(x) e^j(x)\,,$$
\item $\tilde{h}_{u\,v}$ is a symmetric
$(L'-1)\times(L'-1)$ 2-tensor, invariant under $GL(1,\mathbb{R})\,,$
$$\tilde{h}_{uv}[T]y^u(x) y^v(x) =\sum_{\rho \ = irreps\ of\
GL(1,\mathbb{R})}
\tilde{h}^{\rho}[T]\rho_{uv}y^u(x) y^v(x)\,,$$
\item of course, the Einstein-Weyl equations (\ref{ew11}) should also be
imposed,

\item $y^{u_0}$ satisfies the Maurer-Cartan consistency condition :
\beq\label{ewk0} dy^{u_0} = -\frac{1}{2}f_{vw}^{u_0}\,y^v\wedge y^w -
\frac{1}{2}f_{ij}^{u_0}\,\tilde{e}^i\wedge\tilde{e}^j \,.
\eeq
\end{itemize}

\subsection{Some families of solutions}
\begin{itemize}
\item To escape from the no-go theorem of Corollary 1, it may be tempting
to consider a non-semi-simple group $G \equiv GL(1,\mathbb{R})\times
\tilde{G}$ where $(\tilde{G}/H)$ is a (n-2)-dimensional symmetric
space : in that case (note that the unimodularity condition $f_{ij}^j = 0$
is trivially satisfied) there are only two unknown functions of
$T\,:\tilde{h}_0[T]$ and the one that multiplies the unique standard
metric on
$(\tilde{G}/H)\,.$ A particular situation in that family is one,
with a compact group G, considered by Madsen et al
\cite{{Madsen-a},{mpps}} and analysed in Section 5 of the present work~:
there
$G
\equiv S^1\times SO(n-1)\,,\ H \equiv SO(n-2)\,.$ A 4-dimensional non-compact example is the Bianchi VIII case
\cite{Bonneau97} where $ (\tilde{G}/H)
\equiv SU(1,1)/U(1)\,.$

\item Other situations with only two unknown functions of $T$ in
(\ref{ew110}) occur when $L' = 1$ and
$G/(H\times U(1))$ is a compact irreducible symmetric space : this requires
$ {\cal H'}= {\cal U}(1) \simeq {\cal SO}(2) \simeq S^1\,.$ Indeed, this
ensures that the matrix
$h_{ij}[T]$ depends on a single function of
$T$
\footnote{\,The unimodularity condition decomposes into :$f_{ij}^j +
f_{i\,u_0}^{u_0} = 0\,,$ which is true (the indices $i,\,j,\,k$ run among
${\cal (G-H-H')}$ generators), and $f_{u_0\,j}^j = 0$ which results from
the compactness of $H'\,.$}. In that case, the subgroup
$\tilde{H}$ contains a  $U(1)$ factor, and the symmetric space
$G/\tilde{H}$ is  necessarily \cite{Kahlersymmetrique} a K\"ahler space
whose   K\"ahler form $J$ is proportional to the closed 2-form :
\beq\label{ewk1} dy^{u_0} \stackrel{(\ref{ewk0})}{=} -
\frac{1}{2}f_{ij}^{u_0}\tilde{e}^i\wedge\tilde{e}^j = 2 J\,.\eeq   

Let us explicitely prove that the Einstein-Weyl metrics in that family 
are  Riemannian conformally K\"ahler metrics, so generalising our four
dimensional analysis
\cite{Bonneau97}. The metric (\ref{ew110}) writes :
$$  (ds)^2 = (dT)^2  +
\tilde{h}_{0}(T)y^{u_0}\,y^{u_0} +  2\tilde{h_1}(T)
g_{i\,\bar{j}}dz^i\,d\bar{z}^{\bar{j}}$$ and the K\"ahler form $J 
\equiv ig_{i\,\bar{j}}dz^i\wedge d\bar{z}^{\bar{j}}\,.$
 $K(z,\,\bar{z})$ being the K\"ahler potential, the one-form $y^0$ writes 
:
$$ y^0 = dU -i\partial_i K dz^i + i \partial_{\bar{j}} K
d\bar{z}^{\bar{j}}$$ and in the basis $\dst dx^m
\,:\,\left\{dT,\,dU,\,dz^i,\,d\bar{z}^{\bar{j}} \right\}\,,\ 
i,\,\bar{j}\,= 1,2..,(n-2)/2\,,\ m =1,2..,n\,,$ the metric will be written
$(ds)^2 = G_{mp}dx^m dx^p\,.$ Consider now the 2-form
$$\Om = \sqrt{\tilde{h}_0(T)}dT\wedge  y^0 + \tilde{h}_1(T)J\,.$$ In the
basis $\{ dx^m\}\,,$ $\Om = \frac{1}{2} \bar{J}_{mp}dx^m \wedge dx^p$
defines an antisymmetric 2-tensor $\bar{J}_{mp}\,.$ The tensor
$\bar{J}_m^{\ p} = \bar{J}_{mq} G^{qp}$ is found to be 
\beq\label{matrix}
\bar{J}_m^{\ p} = \left[ \begin{array}{cccc}
 0 & \frac{1}{\sqrt{\tilde{h}_0}} & 0 & 0 \\ -\sqrt{\tilde{h}_0} & 0 & 0 &
0 \\ i\sqrt{\tilde{h}_0} \partial_i K & -\partial_i K &i \mathbb{I} & 0 \\
-i\sqrt{\tilde{h}_0} \partial_{\bar{j}} K & -\partial_{\bar{j}}K & 0 &-i
\mathbb{I}  
\end{array}\right] \eeq and ones verifies that $\bar{J}_m^{\ p}
\bar{J}_p^{\ q} = - \de_m^q\,.$ With expression (\ref{matrix}) for 
$\bar{J}_m^{\ p}\,,$ one computes the Nijenhuis tensor and finds it to be
identically zero : we have a complex structure, and, thanks to the
antisymmetry of $\bar{J}_{mp}\,,$ the metric $G_{mp}$ is hermitian with
respect to $\bar{J}\,.$ The differential $d\Om$ is computed and found to be
 $$d\Om = \frac{d\Phi}{dT}  dT\wedge\Om\,,\ {\rm with}\ \frac{d\Phi}{dT} =
\frac{d\log{\tilde{h}_1(T)}}{dT} -
2\frac{\sqrt{\tilde{h}_0(T)}}{\tilde{h}_1(T)}\,,$$ and, after the
conformal transformation, compatible with the cohomogeneity-one property
($\bar{J}_m^{\ p}$ being  unchanged) : $$g \rightarrow \tilde{g} =
g\exp{[-\Phi(T)]}\,,\ \Om \rightarrow \tilde{\Om} = \Om\exp{[-\Phi(T)]} 
\,,\ \ga \rightarrow \tilde{\ga} = \ga  - d\Phi(T)\,, $$ one gets
$d\tilde{\Om} = 0\,.$   As a consequence, we have~:

\vspace{3mm}
\noindent {\bf Theorem 2 :} {\sl Given an arbitrary (n-2)-dimensional
compact symmetric K\"ahler space ${G}/\tilde{H}$ [then 
$\tilde{H} \equiv U(1)\times H$], any  non-exact Einstein-Weyl structure
of cohomogeneity-one under the left action of $G$ has a Riemannian
conformally K\"ahler metric and the principal orbit is the coset
space G/H.}    

Some remarks are in order :\begin{itemize}
\item Note that we only used the cohomogeneity-one structure and the 
existence of an extra Killing vector for Einstein-Weyl structures.
\item The structures are not ``locally conformal K\"{a}hler" ones in the
sense of   Vaisman \cite{{Vaisman},{PS93}} as the complex structure is not
covariantly constant with respect to the Weyl covariant  derivative
$D^{\tilde{\ga}}\,$(this would require $\ga = d\Phi(T)\,).$
\item A particular situation in that family is the one where, in even
dimensions n = 2m,
$G = SU(m)\,,\ H = SU(m -1)\,,\ H' = U(1)\,:$ it was considered by Madsen
et al
\cite{{Madsen-a},{mpps}} and is analysed in Section 4 of the present work
($G/\tilde{H} \equiv \mathbb{C}P^{m -1}$).
\item Another one would be $G = SO(m+1)\,,\ H = SO(m -1)\,,\ H' = SO(2)$
\cite{PS93}, e.t.c....
\end{itemize}
 
\item In other situations, the condition $H' = GL(1,\mathbb{R})\,,$  will
be relaxed, for example in dimensions n = 5 + 4p, with $G/H \equiv SU(p
+2)/SU(p)\,,\ H' = SU(2)\,,$ e.t.c...
\end{itemize}
\vspace{1cm}

\noindent We do not intend to give here a complete classification of
(non-exact) Einstein Weyl structures in an arbitrary dimension, but mainly
to emphasize that the symmetry of Einstein-Weyl solutions is bigger than
that of the equations.


\section{SU(m) invariant structures}  In n = 2m dimensions, the previous
analysis shows that a non-exact  Weyl structure of cohomo geneity-one
under $SU(m)$ has, in a Gauduchon gauge, an extra $U(1)$ invariance, so
extending previous results shown for n = 4
\cite{Madsen-b}. The Weyl structure (\ref{ew110}) may be written :
\beq\label{2-330} ds^2 =  (dT)^2  + f^2(T)(y^0)^2  + h^2(T) g_B\
\ ,\ \ \ga = \pm\Ga f^2(T)y^0\ ,
\eeq where $\Ga$ is a constant positive parameter, $g_B$ is the standard
Fubini-Study metric on
$CP^{m-1}$ with K\"ahler form
$J$, Ricci curvature $ = 2mg_B$ and the one-form $y^0$ is chosen to
satisfy 
$ dy^0 = 2\,J_B$ (\ref{ewk1}),
$i.e.\ \eta =1$ in the notations of \cite{Madsen-a} (Note that Madsen's
parameter $\eta^2 = k^2$ may be reabsorbed into the definition of $\si$ :
all his equations are invariant under the change $f^2 \rightarrow f^2/k^2\
, \be \rightarrow \be/k$ such that
$f^2\si^2$ and $\be\si$ are left unchanged.) Note that here ($dy^0 \neq
0$),  an exact Einstein-Weyl structure requires $\Ga =0\,.$    
\subsection{Local expressions}   The Einstein-Weyl equations (\ref{ew11})
write
\cite{{Madsen-a},{mpps}}:
\beqa\label{2-302b} a)\ \La' & = & -\frac{f''}{f} -(n-2)\frac{h''}{h}
-\frac{1}{2}\Ga^2 f^2\,, 
\nnb\\ c_{(00)})\ \La' & = &  -\frac{f''}{f} -(n-2)\frac{h'f'}{h\,f}
+(n-2)\frac{f^2}{h^4} +\frac{n-4}{4}\Ga^2 f^2\,, 
\\ c_{(ij)})\ \La' & = &    -\frac{h''}{h} -(n-3)\frac{h'^2}{h^2}-
\frac{h'f'}{h\,f} -2\frac{f^2}{h^4} +\frac{n}{h^2} -\frac{1}{2}\Ga^2 f^2
\,.\nnb
\eeqa   To follow as closely as possible our previous 4-dimensional
analysis \footnote{\,There, $(d\tau)^2 =
\si^2_1 +\si^2_2\,,$ where the $\si_i\,,\ i=1,\,2,\,3$ are three $SU(2)$
left-invariant one-forms satisfying $d\si_i =
\frac{1}{2}\epsilon_{ijk}\si_j\wedge
\si_k\,.$} , we rewrite $g_B$ as
$\frac{1}{4}(d\tau)^2\,,\ J_B$ as $\frac{1}{4}J\,,\ y^0$ as 
$\frac{1}{2}\si^3\,,\ d\si^3 = J$ and equation (\ref{2-330}) with
notations inspired by gravitation
\cite{{DS95},{Tod95}} :
\beq\label{2-330bis} ds^2 = \left[\om^2(t)\om_3(t) (dt)^2  +
\frac{\om^2(t)}{\om_3(t)}(\si^3)^2 \right] + \om_3(t)(d\tau)^2\ \ ,\ \
\ga = \pm\Ga\frac{\om^2(t)}{\om_3(t)}\si^3\ .
\eeq

As in \cite{Bonneau97}, define
$u(t)$ through :
\beq\label{2-301} u(t) = \frac{1}{\om_3\,
\om^2}(\frac{d\om_3}{dt} - \om^2)
\ .
\eeq
 The difference of the first two equations (\ref{2-302b}), allows the
calculation of the derivative of $u(t)$ :
$$ \frac{d u}{dt} = -\frac{1}{2}\om^2[\Ga^2 + u^2]\ \ \ <\ 0\,.$$
\noindent Then one can change the variable $t$ into $u$ and compute :
$$ \frac{d \om_3}{du} = - 2\frac{1 + u\om_3}{
\Ga^2 + u^2}\ \
\,,$$ which integrates to :
\beq\label{3-42}
\om_3(u) = 2\frac{k - u}{\Ga^2 + u^2}\,.
\eeq Defining :
\beq\label{3-43}
\Om^2  =  \frac{1}{4}(\Ga^2 + u^2)\om^2 \,,
\eeq and using the Einstein-Weyl equations (\ref{2-302b}), one obtains a
second order linear differential equation  :
\beqa\label{3-44}
\frac{d^2 \Om^2}{d u^2}  -  \left[
\frac{2(m-3)u}{\Ga^2 + u^2} +
\frac{m-4}{k - u}\right]\frac{d \Om^2}{d u} & - &
\left[\frac{3(m-2)(\Ga^2+k^2)}{(k-u)^2}+ m\right]\frac{\Om^2}{\Ga^2 + u^2}
= \nnb \\ & = & -\frac{m}{\Ga^2 + u^2} \,.
\eeqa  Excluding Einstein solutions (as well as exact  Einstein-Weyl
structures), we rescale $u$ and $k$ according to $u = \Ga x\ ,k = \Ga
\ka\ ,$ and get the following ($\Ga$ independant) expression  for
$\Om^2$ :
\beq\label{3-45}
\Om^2(x)  = m\left(\frac{1+x^2}{\ka -x}\right)^{m-2}\left[l_1(x^2-2\ka x -
1) + I_m[\ka,x] -2l_2I_{m+1}[\ka,x]\right]\,,
\eeq where($n \ge 2$) :
\beq\label{3-45b} 
 I_n[\ka,x] = \frac{(\ka -x)^{n-2}}{2[1 + x^2]^{n-2}} +(x^2-2\ka x -
1)\left[(n-2)\int_x^{\ka} \frac{(\ka -y)^{n-3}}{2[1 + y^2]^{n-1}}dy
+\frac{\de_{n,2}}{2(1+ \ka^2)}\right]\,.
\eeq For further use, notice that the functions $I_n[\ka,x]$ may be
expressed as :
\beqa\label{3-45c}
 I_n[\ka,x] & = &  (x^2-2\ka x - 1)J_n(\ka,x) \nnb \\ {\rm with}\
\ \frac{\partial J_n(\ka,x)}{\partial x}\mid_{\ka} & = &
\frac{(\ka -x)^{n-1}}{(x^2-2\ka x - 1)^2[1 + x^2]^{n-2}}\ >\ 0\
\,. \eeqa
\noindent When $x \rightarrow -\,\infty\,,$ the functions
$J_n(\ka,x)$ become 
\beq\label{3-45d}\tilde{J}_n(\ka) =
\de_{n,2}\frac{1}{2(1+\ka^2)} + (n-2)\int_{-\,\infty}^{\ka}
\frac{(\ka -y)^{n-3}}{2[1 + y^2]^{n-1}}dy\ >\ 0\ ,\eeq and one proves that
:
\beq\label{3n-45} J_n(\ka,x) - \tilde{J}_n(\ka) \simeq
\frac{1}{n(-x)^n}\ \,,\ \ x \rightarrow -\,\infty\,.
\eeq
\noindent The behaviour near $\ka$ is \beq\label{3n-45b} J_n(\ka,x)
\simeq -\frac{(\ka-x)^n}{n(1+\ka^2)^n}\ ,\ \ x
\rightarrow \ka^-\,.\eeq Then $J_n(\ka,x)$ is an increasing function from
$\tilde{J}_n(\ka)$ to
$+\,\infty$ when $x$ varies from $-\,\infty$ to $(
\ka-\sqrt{1+\ka^2}\,),$ and from
$-\,\infty$ to zero when $x$ varies from $(\ka-\sqrt{1+\ka^2})$ to
$\ka\,.$ As a consequence,
$I_n(\ka,x)$ is a continuous positive function between
$-\,\infty$ and $\ka$ where it vanishes. These properties will be useful
in the discussion of the regularity of the distance.

  Equations (\ref{3-42},\ref{3-45}) and
\beq\label{3-46}
\frac{du}{dt} =  - 2\Om^2
\eeq give the distance \footnote{\ Of course, the
 parameters $\ka,\ l_1,\ l_2$ and the proper time $x$ are constrained by
positivity  :
$\Om^2 >0\ ,\ \ka -x >0\,.$} and Weyl form as functions of the new
``proper time" $x$ :
\beqa\label{3-47}  ds^2 & = &
\frac{2}{\Ga}\left[\frac{\ka -x}{\Om^2(1 + x^2)^2}(dx)^2
 + \frac{\Om^2}{\ka - x}(\si^3)^2  + 
\frac{\ka - x}{(1 + x^2)}(d\tau)^2 \right]\ ,\nnb \\
\ga & = & \pm\frac{2\Om^2}{\ka -x}\si^3 \ \ .
\eeqa  Finally, the conformal scalar curvature is computed from
(\ref{ew101},\ref{2-302b}) :
\beq\label{3-48}  S^D = m^2 l_2 \Ga -2m(m-2)\Ga\frac{\Om^2}{\ka -x} =
\frac{n\Ga}{4}\left[nl_2 - 2(n-4)\frac{\Om^2}{\ka -x}\right] \le
\frac{n^2\Ga l_2}{4}\ .
\eeq If one looks for solutions with a constant conformal scalar
curvature, equation (\ref{3-44}) can only be satisfied for m = 2
\cite{{Bonneau},{CalderbankFaraday}}.

 As discussed in Subsection (3.2) and Theorem 2, under the conformal
transformation
$\tilde{g} = \Ga [1 +x^2] g/2$, the metric may be rewritten in the
standard form (\ref{2-330bis}) with $$\tilde{\om} =
\sqrt{\Om^2(1 + x^2)}\ ,\ \
\tilde{\om}_3  = \ka - x
\,,$$ the ``proper time" $\tilde{t}$ being given by
$$d\tilde{t} =-\frac{dx}{\Om^2(1 + x^2)}\,.$$  Then,
$$\frac{d\tilde{\om}_3}{d\tilde{t}} - \tilde{\om}^2 = 0\ \ ,$$  ensuring
that the n-dimensional metric $\tilde{g}$ is K\"ahler with K\"ahler form
given by $$ \tilde{J^n} =
\tilde{\om}^2d\tilde{t}\wedge \si^3 + \tilde{\om}_3 J\,,\ \ d\si^3  =
J\,.$$   Then we have proved~: 

\vspace{3mm}
\noindent {\bf Theorem 3}  : {\it The most general 2m dimensionnal
(non-)compact  non-exact Einstein-Weyl structure with isometry $SU(m)\,,\
m\ge 2\,,$  is a 3-parameter structure (plus one homothetic parameter) :
the metric is locally conformally K\"ahler.\nl The conformal scalar
curvature is a constant in the Gauduchon gauge if and only if n= 4
dimensions.}

\vspace{5mm}

In the following Subsection, we shall consider the possible positive
definite and regular $U(m)$ invariant Einstein-Weyl metrics. In his 
Ph.D., Madsen gives a classification of compact solutions. Here, in the
same spirit as in \cite{Bonneau98}, we use the terminology of Gibbons and
Hawking 
\cite{GH} on nuts and bolts, well adapted to the analysis of the
completeness of our candidate metrics on orientable manifolds. We shall
prove that, up to an arbitrary homothetic factor
$\Ga>0$, there exist m+2 one-parameter families of complete Einstein-Weyl
metrics with a non-exact Weyl form, each depending on a {\bf strictly
positive} constant
$l_2$ related to the conformal scalar curvature.

\subsection{Regular metrics}   The function $\Om^2(x)$ has to be positive
on the proper time interval, which is then limited by its zeroes - and let
us recall that positivity also requires $x <
\ka\,.$ So, only four kinds of proper-time interval may occur : $
]-\,\infty\,,\ka[\,,\ [-\,\infty\,, x_0[\,,\ ]x'_0\,,
\ka[\,$ and $]x'_0\,, x_0[\,.$
\nl\noindent The possible singularities of the distance (\ref{3-47}) occur
at
$-\,\infty,\ \ka$ or at a zero of the function $\Om^2(x)\,.$
\begin{itemize}\item {\bf a) Regularity of the distance as $x
\rightarrow -\,\infty\,.$}
\nl\noindent When $x \rightarrow -\,\infty\,, \Om^2(x) \simeq m\de (-x)^m$
where $\de = l_1 +\tilde{J}_m(\ka) -2l_2\tilde{J}_{m+1}(\ka)\,.$ The
behaviour of the distance
  is readily seen to be singular if
$\de \neq 0\,.$ Indeed,
\beq\label{n1}
 ds^2 \sim  \frac{2}{\Ga}\left[\frac{(dx)^2}{m\de (-x)^{(m+3)}} +m\de
(-x)^{m-1}(\si^3)^2
 +
\frac{1}{(-x)}(d\tau)^2 \right]\,,
\eeq
 and the change of variable $\rho = (-x)^{-(m+1)/2}$ leaves a non
removable singularity at $\rho = 0\,.$

Consider now the special case when $\de$ vanishes : thanks to
(\ref{3n-45}), the function
$\Om^2(x)$ goes to $1$ when $x
\rightarrow -\,\infty\,.$ So, the distance behaves as
\beq\label{n11}  ds^2 \sim 
\frac{2}{\Ga}\left[\frac{(dx)^2}{(-x)^3} +\frac{1}{(-x)} [(\si^3)^2 +
(d\tau)^2]
\right]\,,
\eeq  and, after the change
$\rho = (-x)^{-1/2}$:
$$ds^2 \sim \frac{8}{\Ga}\left[(d\rho)^2 +
\frac{\rho^2}{4}[(\si^3)^2 +(d\tau)^2]\right]\
\ ,\
\rho \rightarrow 0\,:$$ the singularity is removable if one choses
cartesian coordinates rather than polar ones. Near the end point $\rho
\rightarrow 0\,,$ the manifold is a point which gives a {\it nut}
\cite{GH}. To sum up, we have~:

\vspace{3mm}
\noindent{\bf Lemma 2 :} {\it if the proper time interval extends down to
$-\,\infty$, the metric can be regular only if $\de \equiv l_1
+\tilde{J}_m(\ka) -2l_2\tilde{J}_{m+1}(\ka) = 0\,,$ and then a nut occurs.}

\vspace{5mm}
\item {\bf b) Regularity of the distance at $x = \ka\,.$}
\nl\noindent Consider now the behaviour of the distance near $x =
\ka$ supposed to be the highest possible value of the proper time
compatible with a positive metric. As
$\Om^2(x) \simeq m(1+\ka^2)^{m-1}(-l_1)(\ka -x)^{-m+2} $, the behaviour of
the distance
  is readily seen to be singular if
$l_1 \neq 0\,.$ Indeed,
\beq\label{n2}
 ds^2 \sim  \frac{2}{\Ga}\left[\frac{(\ka - x)^{m-1}(dx)^2}{-ml_1
(1+\ka^2)^{(m+1)}} - m l_1 (\frac{1+\ka^2}{\ka -x})^{m-1}(\si^3)^2
 + \frac{\ka -x}{1+\ka^2}(d\tau)^2 \right]\,,
\eeq
 and the change of variable $\rho = (\ka-x)^{(m+1)/2}$ leaves a non
removable singularity at $\rho = 0\,.$
\nl\noindent We are left with the case
$l_1 =0\,,$ where, thanks to (\ref{3n-45b}), one finds :
$$\Om^2(x) \simeq
\frac{(\ka -x)^2}{1+\ka^2}\,.$$ One can change the variable $x$ into
$\rho$ given by
$\rho =\sqrt{\ka-x}$ and, here also, get the following  {\it nut} behaviour
for the distance near
$x = \ka \ :$
$$ds^2 \sim \frac{8}{\Ga[1+\ka^2]}\left[(d\rho)^2 +
\frac{\rho^2}{4}[(\si^3)^2 + (d\tau)^2 ]\right]\,,\
\rho \rightarrow 0\,. $$ To sum up, we have~:

\vspace{3mm}
\noindent{\bf Lemma 3 :} {\it if the proper time interval extends up to
$\ka$, the metric can be regular only if $l_1 = 0\,,$ and then a nut
occurs.}

\vspace{5mm}
\item{\bf c)  Regularity of the distance at a zero of
$\Om^2(x)\,.$}

\noindent At last, singularities in the distance may occur at zeroes of
$\Om^2(x)\,.$
\nl\noindent If $\Om^2(x_0) = 0$ with
$\frac{d\Om^2}{dx}(x_0) = 0$ and $x_0 
\neq \ka$, the differential equation (\ref{3-44}) enforces $x_0$ to be a
maximum, which contradicts positivity.
\nl\noindent So, consider the situation with
$\frac{d\Om^2}{dx}(x_0) \neq 0$ and change the variable
$x$ to
$\rho$ according to :
\beq\label{n20b}
 x = x_0 + \rho^2\frac{d\Om^2}{dx}(x_0)\ ;\eeq  using
$\Om^2(x)
\simeq \rho^2[\frac{d\Om^2}{dx}(x_0)]^2\,,$ the distance behaves when
$\rho
\rightarrow 0\ $ as :
\beq\label{n21}    ds^2 \sim
\frac{8(\ka - x_0)}{\Ga[1+x_0^2]^2} \left[(d\rho)^2 +
\rho^2\left((\frac{1+x_0^2}{\ka -
x_0})\frac{d\Om^2}{dx}(x_0)\right)^2(\frac{\si^3}{2})^2 +
 \frac{1+x_0^2}{4}(d\tau)^2\right]\,,
\eeq    and one has~:

\vspace{5mm}\noindent{\bf Lemma 4} : {\it if the function
$\Om^2(x)$ vanishes at $x_0$, the metric can be regular only if
$x_0$ is a bolt of twist p, i.e.}:
\beq\label{n22}
\Om^2(x_0) = 0\ \ ;\ \ \left(\frac{1+x_0^2}{\ka
-x_0}\right)\frac{d\Om^2}{dx}(x_0) =
\pm p\,,\ \ p = 1,\ 2...\eeq Indeed, in such a case, restricting the range
in the angle
$\psi \in [0\,,\ 4\pi\,]$ involved in
$\si^3 = d\psi + \cos\te d\phi\,,$ e.t.c... to the interval
$[0,\ 4\pi/p\,]$, and changing from polar coordinates $(\rho,
\psi/2)$ to cartesian ones, there is no longer a singularity in the
distance when $\rho$ goes to zero. Near the end point $\rho
\rightarrow 0\,, $ the manifold is $\mathbb{C}P^{m-1}\,.$ Moreover, the
$U(1)$ isometry corresponding to changes in $\psi$ becomes
$U(1)/\mathbb{Z}_p\,.$

\noindent If $\frac{d\Om^2}{dx}(x_0) > 0$ the bolt is $+p$ and the proper
time interval extends to $\ka$ or to another bolt
$-p$ at $ x_1 > x_0$; on the other situation, the bolt at
$x_0$ is a $-p$ one and the proper time interval extends down to
$-\,\infty$ or to another bolt $+p$ at $x_2 < x_0\,.$

Condition (\ref{n22}) may be rewritten as a relation between $\ka, l_2$ and
$x_0\,:$
\beq\label{n22b}
\ka -x_0 = (m\pm p)\frac{1+x_0^2}{2(\mp px_0 +m l_2)}\,.\eeq
\end{itemize}
 We have now all the building blocks needed in our discussion on the
regularity of our Einstein-Weyl metrics (note that the positive function 
$\mid\mid \ga\mid\mid^2\ = 2\Ga\,\Om^2(x)/(\ka -x)$ vanishes at both ends
of the allowed ``proper time" intervals).

\subsubsection{Nut-Nut metric} Consider firstly the situation where  the
allowed range of $x$ be the largest one,
$]-\,\infty, \ka\,]\,.$  According to Lemma 3, $l_1 = 0$ for completeness
at $x =
\ka$ ; moreover, a nut at $-\,\infty$  requires (Lemma 2) :
$$ l_1 + \tilde{J}_m(\ka) -2l_2\tilde{J}_{m+1}(\ka) = 0\,.$$
\noindent Hopefully, for a given value of the parameter $l_2$, the
vanishing of
$ f_m(\ka)
\equiv
\tilde{J}_m(\ka) -2l_2\tilde{J}_{m+1}(\ka) \,,$ will determine a unique
value for the parameter
$\ka\,.$\nl\noindent Notice firstly that, thanks to the positivity of the
$\tilde{J}_n\,,$ the function $f_m(\ka)$ can vanish only when
$l_2 > 0\,.$ After an integration by parts, $f_m(\ka)$ may be rewritten as
:
\beq\label{4-n1} f_m(\ka) = -(m-1)\int_{-\,\infty}^{\ka}
\frac{(y+l_2)[\ka -y]^{m-2}}{[1+y^2]^{m}}dy\,.
\eeq
\noindent On the one hand, this function cannot vanish if
$\ka + l_2
\le 0\,.$ On the other hand, after some algebra, one obtains the
differential equation :
\beqa\label{4-n2} (m-2)f_m(\ka) & - & (\ka+l_2)\frac{df_m(\ka)}{d\ka} = A
>0
\nnb \\ A & = & \de_{m,2}\frac{(\ka + l_2)^2}{(1+ \ka^2)^2}
+(m-1)(m-2)\int_{-\,\infty}^{\ka}\frac{(y+l_2)^2[\ka
-y]^{m-3}}{[1+y^2]^{m}}dy\,. \eeqa Then, as the vanishing of
$f_m$ requires $(\ka+l_2) > 0\,,$ the derivative of $f_m$ at any of its
zeroes has to be negative. As a consequence of the continuity of that
function, there exists at most one zero for
$f_m\,.$\nl\noindent Moreover, as the
$\tilde{J}_n(\ka)$ (\ref{3-45d}) are easily seen to satisfy the  recursive
relation  :
\beq\label{4-n3} 4(n-2)\tilde{J}_{n+1}(\ka) = 2\ka
(2n-3)\tilde{J}_{n}(\ka) +(n-1)\tilde{J}_{n-1}(\ka)\ ,\ \ n \ge 3\,,
\eeq their behaviour at infinity may be proven to be 
$$\tilde{J}_{n}(\ka) \simeq \be_n\ka^{n-3}(1 + {\cal O}(1/\ka^2))\ ,\
\be_n =\frac{\pi (2n-5)!}{[2^{n-2}(n-3))!]^2}\,,\ {\rm when}\
\
\ka
\rightarrow +\,\infty\ \ ,\ n \ge3\,,$$ and
  $$\tilde{J}_{n}(\ka) \simeq \de_n(-\ka)^{-n}(1 + {\cal O}(1/\ka^2))\ ,\
\de_n =
\frac{[ (n-1)!]^2}{(2n-2)!}\,,\ {\rm when}\ \ \ka \rightarrow -\,\infty\
\ ,\ n \ge 2\,.$$ As a consequence,
$f_m(\ka)$, positive for $\ka\le -l_2$ and going to $-\,\infty$ when
$\ka \rightarrow +\,\infty$ has one and only one zero : given a positive
number $l_2$, the value of the parameter
$\ka > -l_2$ is uniquely fixed (recall that $l_1 = 0$) and determines, up
to a homothety, one and only one nut-nut metric. 

\subsubsection{Nut-Bolt(1) metric} Consider now the situation where the
range of
$x$ is 
$]-\,\infty, x_0\,]\,,\ \Om^2(x_0) = 0\,,\ x_0 <
\ka\,.$ Thanks to Lemma 2, the nut at $-\,\infty$  requires :
\beq\label{5-n0} l_1 + \tilde{J}_m(\ka) -2l_2\tilde{J}_{m+1}(\ka) =
0\,,\eeq
\noindent and, from Lemma 4 and (\ref{n22b}), we know that a bolt  at
$x_0$ (necessarily a bolt(-1) in order to be compatible with the nut at
the other end) implies the two conditions :
\beq\label{5-n1}
\ka = x_0 +(m-1)\frac{1+x_0^2}{2(x_0 + ml_2)}\ \ ; \ \ l_1(x_0^2-2\ka x_0
- 1) + I_m[\ka,x_0] -2l_2I_{m+1}[\ka,x_0] = 0\,.
\eeq Hopefully, for a given value of the parameter $l_2\,,$ these
equations will determine uniquely the other ones ($\ka\,,l_1$) and fix the
metric (\ref{3-47}). From (\ref{5-n0},\ref{5-n1}) one gets the condition :
\beq\label{5-n2} g_m[\ka,x_0] \equiv [J_m(\ka,x_0) -
\tilde{J}_m(\ka)] -2l_2[J_{m+1}(\ka,x_0) -
\tilde{J}_{m+1}(\ka)] = 0\,\eeq Thanks to the increasing character of the
function
$J_n[\ka,x]$ as a function of $x \ < \ka-\sqrt{1+\ka^2}$, both square
brackets in that equation are positive in that range for
$x\,.$ On the other hand, when
$\ka-\sqrt{1+\ka^2} <x \le \ka\,,$ they are both negative. Then, the
existence of a solution again requires a strictly positive
$l_2\,.$

After an integration by parts, $g_m[\ka,x_0]$ may be rewritten as :
\beq\label{5-n4}  g_m[\ka,x_0] = (m-1)\int_{-\,\infty}^{x_0}
\frac{(y +l_2)[\ka -y]^{m-2}}{[1+y^2]^{m}}dy +
\frac{(m-1)}{2m(1+x_0^2)}\left(\frac{m-1}{2(x_0 +m l_2)}\right)^{m-2}\,.
\eeq

\noindent The $J_n(\ka,x)$ satisfy the following recursion relation ($n
\ge 3$) : 
\beq\label{5-n5}  4(n-2)J_{n+1}(\ka,x) = 2\ka (2n-3)J_{n}(\ka,x)
+(n-1)J_{n-1}(\ka,x) -
\frac{(\ka-x)^{n-1}}{(x^2 -2\ka x -1)[1+x^2]^{n-2}}\,,
\eeq  and the same is true for the, positive, square bracket
$[J_n(\ka,x) -
\tilde{J}_n(\ka)]\,.$
\noindent As $\ka$ and $x_0$ are related variables, and as from
(\ref{5-n1}) $\ka -x_0 > 0$ needs $x_0 > -ml_2\,,\ x_0
\rightarrow -ml_2^+\,$ implies $\ka \rightarrow +\,\infty\,,$ and the
behaviour of
$[J_n(\ka,x) -
\tilde{J}_n(\ka)]$ at infinity may be shown to be~: 
$$[J_n(\ka,x) -
\tilde{J}_n(\ka)] \simeq \ga_n\ka^{n-3}(1 + {\cal O}(1/\ka))\ \ {\rm for\
some\ positive} \
\ga_n > 0\,, \ \ n \ge 2\,.$$ 
\noindent As a consequence,
$g_m[\ka,x_0] \simeq -2l_2 \ga_{m+1}\ka^{m-2}$, goes to
$-\,\infty$ when {$x_0
\rightarrow -ml_2\,,\
\ka
\rightarrow +\,\infty\,.$} \nl In the same manner, when $x_0
\rightarrow +\,\infty\,,\ i.\,e.\ \ka
\simeq (m+1)x_0/2
\rightarrow +\,\infty\,,$ one can prove that $g_m[\ka,x_0] \simeq -2l_2
(-\be_{m+1}\ka^{m-2})\,.$ Then it goes to
$+\,\infty$ when {$x_0
\rightarrow\ +\,\infty\,,\
\ka
\rightarrow +\,\infty\,.$}
  
To sum up,
$g_m(\ka(x_0),x_0)$, varying continuously from $-\,\infty$ to
$+\,\infty$ when $x_0$ grows from $-ml_2$ to
$ +\,\infty\,,$ has {\bf at least one zero}. We do not succeed in proving
that the solution is unique, but our previous results for n = 2m = 4
\cite{Bonneau98} and computer analysis of the function
$g_m[\ka(x_0),\,x_0]$ defined through equations (\ref{5-n4},\ref{5-n1})
made us confident on the fact that the parameter
$\ka > -l_2$ {\bf is uniquely fixed} and, due to (\ref{5-n0}), so is
$l_1\,.$ Finally, given a positive parameter $l_2$, there is one and only
one nut-bolt Einstein-Weyl regular metric.

\subsubsection{Bolt(1)-Nut metric} Consider now the situation where the
range of
$x$ is 
$ [x'_0,\,\ka\,]\,,\ \Om^2(x'_0) = 0\,,\ x'_0 < \ka\,.$  From Lemma 3, the
nut at $\ka$ requires $l_1 = 0\,,$ and, from Lemma 4 and (\ref{n22b}), we
know that a bolt  at
$x'_0$ (necessarily a bolt(+1) in order to be compatible with the nut at
the other end) implies the two conditions :
\beq\label{6-n1}
\ka = x'_0 +(m+1)\frac{1+x_0^{'\,2}}{2(-x'_0 + ml_2)}\ \ ; \ \
l_1(x_0^{'\,2}-2\ka x'_0 - 1) + I_m[\ka,x'_0] -2l_2I_{m+1}[\ka,x'_0] = 0\,.
\eeq The same manipulations as in the previous subsection lead to a
similar function :
\beqa\label{6-n2} g'_m[\ka,x'_0] & \equiv & [J_m(\ka,x'_0) -
\tilde{J}_m(\ka)] -2l_2[J_{m+1}(\ka,x'_0) -
\tilde{J}_{m+1}(\ka)] =\nnb\\ & = &  -(m-1)\int_{x'_0}^{\ka}
\frac{(y +l_2)[\ka -y]^{m-2}}{[1+y^2]^{m}}dy +
\frac{(m+1)}{2m(1+x_0^{'\,2})}\left(\frac{m+1}{2(-x'_0 +m
l_2)}\right)^{m-2}\,.\eeqa

\noindent With 
\beq\label{6-n3} x_0 = - \frac{(m-1)x'_0 + 2ml_2}{(m+1)}\,,
\eeq  $\ka(x'_0)$ of equ.(\ref{6-n1}) expressed as a function of
$x_0\,,$ has exactly the same value as
$\ka(x_0)$ of (\ref{5-n1}). Under the same change of variable, it is shown
in Appendix B that the function
$g'_m[\ka, x'_0]$ becomes $-g_m[\ka,x_0]$ of the previous subsection
(\ref{5-n4}). Then, except the different values of
$l_1\,,$ the metrics are the same [as discussed for $m = 2$  in
\cite{Bonneau98}, only the orientation of the Einstein-Weyl manifold
changes.]

\subsubsection{Bolt(p)-Bolt(p) metric}  Consider finally the situation
where the range of
$x$ is 
$[x'_0,\,x_0]\,,\ \Om^2(x_0) = 0\,,\ \Om^2(x'_0) = 0\,,\ x'_0 < x_0 <
\ka\,.$ From Lemma 4 and (\ref{n22b}), we know that a bolt(+p)  at
$x'_0$ and a bolt(-p) at $x_0$ imply four relations between
$\ka,\ x_0,\ x'_0,\ l_1$ and $l_2\,:$ 
\beqa\label{7-n1}
\ka =  x'_0 +(m+p)\frac{1+x_0^{'\,2}}{2(-px'_0 + ml_2)} & = & x_0
+(m-p)\frac{1+x_0^2}{2(px_0 + ml_2)}\nnb \\ l_1(x_0^{'\,2}-2\ka x'_0 - 1) 
+  I_m[\ka,x'_0] -2l_2I_{m+1}[\ka,x'_0] & = & 0\\ l_1(x_0^2-2\ka x_0 - 1)
+ I_m[\ka,x_0] -2l_2I_{m+1}[\ka,x_0]& = & 0\,.\nnb\eeqa The first two
equations, giving a second order algebraic equation for $x_0\,,$ lead to :
\beq\label{7-n2} {\rm solution\ a)}\  :\ x_0 = -\frac{(m-p)x'_0
+2ml_2}{(m+p)}\ ;\ {\rm solution\ b)}\ :\ x_0 =
\frac{(ml_2x'_0 +p)}{(-px'_0 + ml_2)}\,.
\eeq The two others, after operations similar to the ones done in the two
previous subsections,
 lead to the vanishing of a new function :
\beqa\label{7-n3} h_m[x_0,x'_0] & \equiv & [J_m(\ka,x_0) - J_m(\ka,x'_0)]
-2l_2[J_{m+1}(\ka,x_0) - J_{m+1}(\ka,x'_0) ] =\nnb\\  & =  &
(m-1)\int_{x'_0}^{x_0}
\frac{(y +l_2)[\ka -y]^{m-2}}{[1+y^2]^{m}}dy  + 
\frac{(m-p)}{2m(1+x_0^2)}\left(\frac{m-p}{2(px_0
+ml_2)}\right)^{m-2}\nnb\\ & -
&\frac{(m+p)}{2m(1+x_0^{'\,2})}\left(\frac{m+p}{2(-px'_0
+ml_2)}\right)^{m-2}\,.\eeqa

\noindent Of course, here also one finds no solution when $l_2
\le 0\,.$ We shall first prove that if $p \ge m\,,$ there is no solution,
second that for $p < m\,$ relation (\ref{7-n2}-b)) is excluded.
\begin{itemize}

\item {\it Case $p = m\,.$}\nl\noindent As $x_0 \neq \ka$, equation
(\ref{7-n1}) enforces
$x_0 = -l_2\,,$ and $x'_0$ is related to $\ka$ by $\dst\ka =\frac{l_2 x'_0
+1}{l_2 - x'_0}\,.$ The function
$h_m[-l_2,\,x'_0]$ reduces itself to the sum of two strictly negative
terms (the quotient
$\dst\frac{m-p}{2(px_0 + ml_2)} \equiv \dst\frac{\ka -x_0}{1+x_0^2} =
\dst\frac{1}{l_2 -x'_0}$ being finite as $x'_0 < x_0 = -l_2\,$). So there
is no Bolt(m)-Bolt(m) Einstein-Weyl metric.
\item {\it Case $p > m\,.$}\nl\noindent One  has $x'_0 < x_0 < -ml_2/p
\,.$ This condition is readily seen to contradict solution (\ref{7-n2}-b))
and one is left with solution a). Then, the positivity of $x_0 - x'_0 =
-2m\dst\frac{x'_0 + l_2}{p-m}$ enforces
$x'_0 < -l_2\,,$ and the relation $x_0 + l_2 =
\dst\frac{p-m}{p+m}(x'_0 +l_2)$ also ensures that
$x_0 <-l_2\,.$ As a consequence, the function $h_m[x_0,\,x'_0]$ reduces
itself to the sum of three strictly negative terms and there are no
Bolt(p)-Bolt(p) Einstein-Weyl metric for $p > m\,.$  Then we have~:

\vspace{3mm}
\noindent{\bf Lemma 5 :} {\it Regular bolt-bolt Einstein-Weyl SU(m)
invariant metrics, non conformally Einstein, may exist only with a twist
$p < m\,.$}

\vspace{3mm}

Note that this was only conjectured in \cite{Madsen-b}.

\item {\it Case $p < m\,.$} \nl\noindent Consider first the candidate
solution (\ref{7-n2}-b)). Using relations (\ref{7-n1}) and some identities
:
$$\frac{px_0 +ml_2}{1 +x_0^2} = \frac{-px'_0 +ml_2}{1 + x_0^{'\,2}}\ ,\
\ 1 + x_0 x'_0 =ml_2\frac{1 + x_0^2}{px_0 + ml_2}\ ,\ \ (\ka - x_0) + (\ka
- x'_0) = m\frac{1 + x_0^2}{px_0 + ml_2}\,,$$
\noindent one may rewrite the function $h_m$ as :
\beq\label{7-n4}
 h_m[x_0,x'_0] =  -\left(\frac{m-1}{m}\right)\left(\frac{px_0 +
ml_2}{1+x_0^2}\right)\int_{x'_0}^{x_0}
\frac{(y -x'_0)(x_0 -y)[\ka -y]^{m-2}}{[1+y^2]^{m}}dy\,,\eeq whose
negatively definite property ensures that there is no ``solution b)"
candidate.

\end{itemize}
\noindent Then one is left with $p < m$ and the linear relation a)
$x_0 = -\dst\frac{(m-p)x'_0 +2ml_2}{(m+p)}$ between $x_0$ and
$x'_0\,,$
\noindent Some useful identities result from the previous relation : $$0 <
x_0 - x'_0 = -\frac{2m}{m+p}(x'_0+l_2) =
\frac{2m}{m-p}(x_0+l_2) \,,$$ and imply $$x_0 > -l_2 > - ml_2/p\ ;\
\ x'_0 < - l_2 < ml_2/p\,.$$ One then obtains the $x'_0
\rightarrow -\,\infty\,,\ \ x_0 \rightarrow +\,\infty$ limit of
$h_m[x_0,\,x'_0]$ to be $+\,\infty\,:$
\beq\label{7-n5} h_m[x_0,\,x'_0] \simeq -2l_2[-2\be_{m+1}(\ka)^{m-2}]\ \
,\ \
 x'_0 \rightarrow -\,\infty\,,\ \ x_0 \rightarrow +\,\infty\,.\eeq With
regards to the limit $x'_0 \rightarrow -l_2\,,\ \ x_0
\rightarrow -l_2\,,$ one gets 
\beq\label{7-n6} h_m[-l_2,\,-l_2] = -\frac{p}{m(1+l_2^2)^2(2l_2)^{m-2}} <
0\,.\eeq Then there exists {\bf at least one zero} of
$h_m[x_0(x'_0),\,x'_0]\,.$  We do not succeed
 in proving that the solution is unique, but our previous results for n =
2m =4 \cite{Bonneau98} and computer analysis of the function
$h_m[x_0(x'_0),\,x'_0]$ defined through equation (\ref{7-n3}) made us
confident that the parameter
$\ka > -l_2$ is uniquely fixed and, due to (\ref{7-n1}), so is
$l_1\,.$ Finally, given a positive parameter $l_2$, there is one and only
one bolt-bolt Einstein-Weyl regular metric, and we have~:

\vspace{3mm}
\noindent{\bf Lemma 6 :} {\it Regular bolt-bolt Einstein-Weyl SU(m)
invariant metrics, non conformally Einstein, exist for any twist
$p < m\,,$ and depend on a single positive parameter $l_2\,.$}

\vspace{3mm}
\noindent Note also that relation (\ref{7-n2}-a) implies :
\beq\label{7-n2a}
\frac{\ka - x_0}{1+x_0^2} = \frac{\ka - x'_0}{1+x_0^{'\,2}}\quad
\Leftrightarrow \quad \om_3[x_0] = \om_3[x'_0] \,.\eeq

\subsection{Summary} In our Gauduchon gauge, we found m+2, and only m+2,
families of non-conformally Einstein regular
 Einstein-Weyl SU(m) invariant metrics ;  according to the classification
of Gibbons and Hawking, they are complete and live on a compact orientable
manifold without boundary.

 The same analysis with two functions of T  could have been done for any
other (n-2)-dimensional symmetric K\"ahler space with little changes, for
example for the Grassmannian
$SU(p+q)/(SU(p)\times SU(q)
\times U(1))\,,\quad {\rm with}\ pq =(n-2)/2\,.$


\section{$S^1\times SO(n-1)$ invariant structures} Cohomogeneity-one Weyl
structures (\ref{ew110}) with $S^1\times SO(n-1)$  invariance may be
written in a Gauduchon gauge as (here, thanks to (\ref{ewk0}), $dy^0 = 0
\Rightarrow y^0 = d\te\,)$
\cite{Madsen-a} :
\beq\label{2-430} ds^2 =  (dT)^2  + f^2(T)(d\te)^2  + h^2(T) g_B\
\ ,\ \ \ga =
\pm\Ga f^2(T) d\te\ ,\ \te \in (0,\,2\pi)\,,
\eeq where $g_B$ is the standard metric on $S^{n-2}\,,$ with Ricci
curvature $= (n-3)g_B\,.$ Note that an exact structure exists iff. $
f^2(T) =$ constant. 
\subsection{Local expressions} The Einstein-Weyl equations (\ref{ew11})
write
\cite{{Madsen-a},{mpps}}:
\beqa\label{2-402b} a)\ \La' & = & -\frac{f''}{f} -(n-2)\frac{h''}{h}
-\frac{1}{2}\Ga^2 f^2\,, 
\nnb\\ c_{(00)})\ \La' & = &  -\frac{f''}{f} -(n-2)\frac{h'f'}{h\,f}
 +\frac{n-4}{4}\Ga^2 f^2\,, 
\\ c_{(ij)})\ \La' & = &    -\frac{h''}{h} -(n-3)\frac{h'^2}{h^2}-
\frac{h'f'}{h\,f} +\frac{n-3}{h^2} -\frac{1}{2}\Ga^2 f^2
\,.\nnb
\eeqa  Note that an exact structure solution exists and writes :
$$ds^2 = \frac{4f^2}{\Ga^2}\left[(dt')^2  + \sin^2{t'} g_B
+\frac{\Ga^2}{4}(d\te)^2 \right] \ ,\ \
\ga = \pm\Ga f^2 d\te\ ;$$ the metric is the standard metric on $S^1\times
S^{n-1}\,.$ 

Here again, we rewrite (\ref{2-430}) with notations inspired by gravitation
\cite{{DS95},{Tod95}} :
\beq\label{2-430bis} ds^2 = \left[\om^2(t)\om_3(t) (dt)^2  +
\frac{\om^2(t)}{\om_3(t)}(d\te)^2 \right] + \om_3(t)(d\tau)^2\ \ ,\ \
\ga = \pm\Ga\frac{\om^2(t)}{\om_3(t)}d\te\,,
\eeq and define $u(t)$ through :
\beq\label{2-401} u(t) = \frac{1}{\om_3\,
\om^2}\frac{d\om_3}{dt}
\ .
\eeq The difference of the two first equations (\ref{2-402b}), allows the
calculation of the derivative of $u(t)$ which is found to have the same
expression as in Subsection 4.1 :
$$ \frac{d u}{dt} = -\frac{1}{2}\om^2[\Ga^2 + u^2]\ \ \ <\ 0\,.$$
\noindent Then one can change the variable $t$ into $u$ and compute :
$$ \frac{d \om_3}{du} = - 2\frac{ u\om_3}{
\Ga^2 + u^2}\ \
\,,$$ which integrates to :
\beq\label{3-442}
\om_3(u) = \frac{2k}{(\Ga^2 + u^2)}\ \ ,\ k > 0\ \ {\rm thanks\ to\
positivity}.
\eeq Defining :
\beq\label{3-443}
\Om^2  =  \frac{1}{4}(\Ga^2 + u^2)\om^2\,,
\eeq and using the Einstein-Weyl equations (\ref{2-402b}), after a
rescaling of
$u$ and $k$ according to
$u =
\Ga x\ ,k = \Ga \ka$, one obtains a second order linear differential
equation  :
\beq\label{3-444} (1+x^2)\frac{d^2 \Om^2}{d x^2}  -  (n-6)x\frac{d
\Om^2}{dx} - 2(n-3)[\Om^2  - 1] = 0\,.
\eeq  It solves to :
\beq\label{3-445}
\Om^2(x)  = 1 - l_1 x(1+x^2)^{(n-4)/2}  - l_2 [1 + (n-3)x(1
+x^2)^{(n-4)/2} K_n(x)]\,,
\eeq  where ($n \ge 3$)  :
\beq\label{3-445b} K_n(x) = \int_0^x
\frac{dy}{(1+y^2)^{(n-2)/2}}\,.
\eeq For further use, notice that when $x \rightarrow
\pm\,\infty\,,$ the functions
$K_n(x)$ behave as ($n \ge 4$) :
\beq\label{3-445d} K_n(x) \simeq \pm[a_n +\frac{1}{(n-3)\mid x\mid^{n-3}}]
\,,\ \ x \rightarrow
\pm\,\infty\ ;\ a_{n} =
\frac{\Ga[(n-3)/2]\Ga[1/2]}{2\Ga[(n-2)/2]}\ .
\eeq

  \noindent Equations (\ref{3-442},\ref{3-445}) and
\beq\label{3-446}
\frac{du}{dt} =  - 2\Om^2
\eeq give the distance \footnote{\ Of course, the
 parameters $\ka,\ l_1,\ l_2$ and the proper time $x$ are constrained by
positivity  :
$\Om^2 >0\ ,\ \ka >0\,.$} and Weyl form as functions of the new ``proper
time" $x$ :
\beqa\label{3-447}  ds^2 & = &
\frac{2\ka}{\Ga}\left[\frac{(dx)^2}{\Om^2(x)(1 + x^2)^2}
 + \frac{\Om^2(x)}{\ka^2}(d\te)^2  + 
\frac{(d\tau)^2}{(1 + x^2)} \right]\ ,\nnb \\
\ga & = & \pm\frac{2\Om^2(x)}{\ka}d\te\ .
\eeqa  For further reference, note that the positive parameter
$\ka$ only appears in the combination $d\te/\ka\,,$ and as a rescaling of
the homothety parameter $\Ga\,.$

The distance may be rewritten as a function of the angle
$\Psi \in [0,
\pi]\,,\cot\Psi = x$
\beqa\label{3-447b}  ds^2 & = &
\frac{2\ka}{\Ga}\left[\frac{(d\Psi)^2}{\Om^2(\Psi)}
 + \frac{\Om^2(\Psi)}{\ka^2}(d\te)^2  + 
\sin^2 \Psi(d\tau)^2 \right]\ ;\\ 
\Om^2(\Psi) & = & 1-l_2 -\cos\Psi \sin^{3-n}\Psi\left[l_1 +(n-3)l_2
\int_{\Psi}^{\pi/2} \sin^{n-4}\phi d\phi\right]\ .\nnb\eeqa Notice that for
n = 3, the differential equation (\ref{3-444}) solves to $\Om^2 = 1 - l_2 -
l_1\cos\Psi\,,$ in agreement with (\ref{3-447b}) :  $\Om^2(\Psi)$ varies
monotonically between
$1-l_1-l_2$ and $1+l_1 -l_2\,,$ then it has  at most one zero.

 Finally, the conformal scalar curvature is computed from
(\ref{ew101},\ref{2-402b}) :
\beq\label{3-448}  S^D  = \frac{\Ga}{2\ka}\left[n l_2 + n(n-4)(1
-\Om^2(x))\right]\ \le \frac{\Ga }{2\ka}n(l_2+n-4)\ .
\eeq Note that a constant conformal scalar curvature requires either n = 4
or
$\Om^2(x) = 1\,.$ This last case corresponds to an exact Weyl form (note
that in our local approach, a closed Weyl-form is an exact one) and the
metric (\ref{3-447b}) is the standard metric on $S^1\times S^{n-1}$
\cite{CalderbankFaraday}.  Then we have proved \footnote{\ For n = 3, the
ansatz (\ref{2-430}) coresponds to the special case $f = 0$ of Tod's
general analysis on 3 dimensional Einstein-Weyl structures
\cite{Tod92} : his 4 parameters ($f\,,\ \la\,,\ B\,$ and
$C$) may respectively be rewritten as
$f = 0\,,\ \la = \Ga\,,\ B = (1-l_2)\Ga/(4\ka)\,$ and $ C = [(l_1)^2
-(1-l_2)^2]/(4\ka^2)\,;$ his coordinates are respectively $V =
\sqrt{2\Om^2(x)/(\ka\Ga)}\,,\ t = \te\,$ and $ (dy) =
\sqrt{8\ka^3/(\ga\,l_1^2)}(d\tau)\,.$}~:

\vspace{3mm}
\noindent {\bf Theorem 4}  : {\it The most general n $\ge 4$ dimensional
(non-)compact  non-exact Einstein-Weyl structure with a
$S^1\times SO(n-1)$ invariant metric is  a 3-parameter structure (plus one
homothetic parameter).\nl The metric has a constant conformal curvature in
the Gauduchon gauge if and only if the dimension n = 4.}

\vspace{5mm} In the following Subsection, we shall consider the possible
positive definite and regular $S^1\times SO(n-1)$ invariant Einstein-Weyl
metrics, still with the tools of nuts and bolts. We shall prove that, up
to an arbitrary homothetic factor
$\Ga>0$, there exist three one-parameter families of complete
Einstein-Weyl metrics with a non-exact Weyl form, depending on a {\bf
strictly positive} constant
$l_2$ related to the conformal scalar curvature.

\subsection{Regular metrics}   The function $\Om^2(x)$ has to be positive
on the proper time interval. The possible singularities of the distance
occur at $x =\pm\,\infty,$ or at a zero of the function
$\Om^2(x)\,.$ The case n = 3, which requires a special analysis as the
candidates are not solely given by the ansatz (\ref{2-430})
\cite{Tod92}, will not be considered in the following.
\begin{itemize}\item {\bf a) Regularity of the distance as $x
\rightarrow
\pm\,\infty\,.$}
\nl\noindent When $x \rightarrow \pm\,\infty\,,\ \Om^2(x) \simeq
-\de_n^{\pm} \mid x\mid^{n-3}$ where
$\de_n^{\pm} = l_1 \pm (n-3)a_n l_2\,.$ As above, the behaviour of the
distance
  is readily seen to be singular if
$\de_n^{\pm} \neq 0\,.$ 

Consider now the special cases when $\de_n^{\pm}$ vanishes : thanks to
(\ref{3-445d}), the function
$\Om^2(x)$ goes to $1$ when $x
\rightarrow \pm\,\infty\,.$ So, the distance behaves as
\beq\label{n411}  ds^2 \sim 
\frac{2\ka}{\Ga}\left[\frac{(dx)^2}{(x)^4} +\frac{1}{\ka^2}(d\te)^2 +
\frac{1}{x^2}(d\tau)^2
\right]\,.
\eeq  Under the  change
$\rho = 1/x$:
$$ds^2 \simeq \frac{2\ka}{\Ga}\left[(d\rho)^2 + 
\rho^2 (d\tau)^2 + \frac{1}{\ka^2}(d\te)^2 \right]\
\ ,\
\rho \rightarrow 0\,.$$ The singularity is removable if one changes to
cartesian coordinates in the (n-1) dimensional space : near the end point
$\rho = 0\,,$ the manifold is a circle $S^1$ which, generalizing Gibbons
and Hawking terminology \cite{GH}, we call a {\it bolt $(S^1)$ }. To sum
up, we have~:

\vspace{3mm}

\noindent{\bf Lemma 7 :} {\it if the proper time interval extends to
$\pm\,\infty$, the metric can be regular only if $\de_n^{\pm}
\equiv  l_1 + l_2 (n-3)K_n(\pm\,\infty) = 0\,,$ and then a bolt
$(S^1)$ occurs.}\nl\noindent A Corollary is that the sole solution with
$]-\,\infty\,,\ +\,\infty\,[$ as proper time interval, requires
$l_1 = l_2 =0\ i.e. \
\Om^2(x) =1\,$ which leads to the metric on the $S^{n-1}$ sphere.

\vspace{3mm}

\item{\bf b)  Regularity of the distance at a zero of
$\Om^2(x)\,.$}

If $\Om^2(x_0) = 0$ with $\frac{d\Om^2}{dx}(x_0) = 0\,,$ the differential
equation (\ref{3-444}) enforces $x_0$ to be a maximum, which  contradicts
positivity. Then, change the variable
$x$ to
$\rho$ according to :
\beq\label{n420b}
 x = x_0 + \rho^2\frac{d\Om^2}{dx}(x_0)\ ;\eeq  using
$\Om^2(x)
\simeq \rho^2[\frac{d\Om^2}{dx}(x_0)]^2\,,$ the distance behaves when
$\rho
\rightarrow 0\ $ as :
\beq\label{n421}    ds^2 \simeq
\frac{8\ka}{\Ga[1+x_0^2]^2} \left[(d\rho)^2 +
\rho^2
\left((\frac{1+x_0^2}{2\ka})\frac{d\Om^2}{dx}(x_0)\right)^2(d\te)^2 +
 \frac{1+x_0^2}{4}(d\tau)^2\right]\ ,\
\rho \rightarrow 0\,.
\eeq If $$\frac{(1+x_0^2)}{2\ka}\frac{d\Om^2}{dx}(x_0) = \pm p\,,\
\ p =1,2,..$$
\noindent the singularity is removable if one changes to cartesian
coordinates in the 2 dimensional space ($\rho\,,\
\te$), and restricts the range in the angle
$\te$ to the interval
$[0,\ 2\pi/p ]$ : near the end point
$\rho = 0\,,$ the manifold is the sphere $S^{n-2}$ which gives a {\it bolt
}
\cite{GH}. As was previously remarked, the integer $p$ that "divide" the
$\te$ interval, may be  reabsorbed into the definition of the parameters
$\ka$ and $\Ga\,:$ so, without loss of generality, we shall only consider
$p = 1\,.$ To sum up, we have~:

\vspace{3mm}
\noindent{\bf Lemma 8 :} {\it if the function $\Om^2(x)$ vanishes at
$x_0$, the metric can be regular only if $x_0$ is a bolt($S^{n-2})$ of
twist 1, i.e.}:
\beq\label{n422}
\Om^2(x_0) = 0\ \ ;\ \
\left(\frac{1+x_0^2}{2\ka}\right)\frac{d\Om^2}{dx}(x_0) =
\pm 1\,,\eeq

\noindent If $\frac{d\Om^2}{dx}(x_0) > 0$ the bolt is $+1$ and the proper
time interval extends up to $+\,\infty$ or to another bolt
$-1$ at $ x_1 > x_0$; on the other situation, the bolt at
$x_0$ is a $-1$ one and the proper time interval extends down to
$-\,\infty$ or to another bolt $+1$ at $x_2 < x_0\,.$

Condition (\ref{n422}) may be rewritten as a relation between
$\ka, l_2$ and $x_0\,:$
\beqa\label{n422b}
\ka &  = & \mp\frac{1 - l_2 +(n-3)x_0^2}{2 x_0}\,,\ {\rm if}\ x_0
\neq 0\nnb\\
\ka & = & \frac{(n-3)a_n}{2}\,,\ {\rm if}\ x_0 = 0
\Leftrightarrow l_2 = 1\,.
\eeqa
\end{itemize}
 \noindent We have now all the building blocks needed in our discussion on
the regularity of our Einstein-Weyl metrics, according to the possible
proper time intervals.

\subsubsection{Bolt($S^1)$-Bolt($S^1)$ metric} Consider a situation where
the allowed range of $x$ is the largest one
$]-\,\infty,\,+\infty\,[\,.$  According to Lemma 7, $\Om^2(x) = 1$  and
the Einstein-Weyl structure is {\bf an exact one}, conformal to the
Einstein case $S^1 \times S^{n-1}\,.$ Then we are not interested.

\subsubsection{Bolt($S^{n-2})$-Bolt($S^1)$ metric} Consider now a situation
where the allowed range of
$x$ is $[\,x_0\,,\,+\,\infty\,[\,,$  with $\Om^2(x_0) = 0\,.$  Thanks to
Lemma 7, $$l_1 = -(n-3)a_n l_2$$ and, from Lemma 8 and (\ref{n422b}), we
know that a bolt $(S^{n-2})$ at
$x_0$  implies the two conditions :
\beqa\label{5-4n1} x_0 \neq 0\ :\ \ka = -\frac{1 -l_2 +(n-3)x_0^2}{2x_0}\
& \quad {\rm or} &  x_0 = 0 \Leftrightarrow l_2 = 1\ :\
\ka =
\frac{(n-3)a_n}{2}\ ; \nnb \\
\Om^2(x_0) &  = & 0\,.
\eeqa The derivative of $\Om^2$ may be written as :

\beqa\label{5-4n1c}
\frac{d\Om^2}{dx} & = & (n-3)l_2[1+(n-3)x^2](1+x^2)^{(n-6)/2} G(x)\,,\nnb
\\
 G(x) & = & -\frac{x}{[1+(n-3)x^2](1+x^2)^{(n-4)/2}}
+\int_x^{\infty}\frac{dy}{(1+y^2)^{(n-2)/2}} \,,\\
 \frac{dG}{dx} & = & -\frac{2(1+x^2)}{[1+(n-3)x^2]^2(1+x^2)^{(n-2)/2}}\qq\
< 0 \,.\nnb \eeqa 
$G(x)\,,$ decreasing from $2a_n$ to $0$ between $x = -\,\infty$
 and $x = +\,\infty\,,$ is strictly positive ($G(0) = a_n).$ As a
consequence, if
$l_2 \le 0\,,$  $\Om^2(x)$ monotonicaly decreases from
$+\,\infty$ to $1$ and cannot vanish. On the contrary, if $l_2 > 0\,,$
$\Om^2(x)$  monotonicaly increases from
$-\,\infty$ to $1$ and  its vanishing determines  a unique value for the
parameter
$x_0\,.$\nl Notice also that in the range
$[\,x_0\,,\,+\,\infty\,[\,,$
 \beq\label{t-3} 0 \le \Om^2(x) < 1\ \Rightarrow l_2\frac{n\Ga}{2\ka}
\le S^D \le (l_2 +(n-4))\frac{n\Ga}{2\ka}\,:\eeq  the conformal scalar
curvature is a strictly positive function on the manifold, whatever the
dimension $n \ge 4\,,$ be, in agreement with a theorem of Calderbank for
the compact case
\cite{CalderbankFaraday}.

 To summarize, given a positive parameter $l_2$, $l_1$ and $\ka$ are
fixed, and, up to an homothethy, there is one and only one
$S^{n-2}-S^1$ Einstein-Weyl regular metric. Its scalar conformal curvature
is a stricly positive function on the manifold.

The particular case $l_2 = 1$ requires $x_0 = 0\,,\ l_1 = -(n-3)a_n$ and
$\ka = (n-3)a_n/2\,.$

\subsubsection{Bolt($S^1)$-Bolt($S^{n-2})$ metric} Consider now a situation
where the allowed range of
$x$ is
$]\,-\infty\,,\ x'_0\,]$  with $\Om^2(x'_0) = 0\,.$ The same discussion as
in the previous subsection ($\Om^2(x)$ is unchanged when $x
\rightarrow -x$ and $l_1 \rightarrow -l_1$) gives a unique solution for
$x'_0$ for any
$l_2 > 0\,$($x'_0 = - x_0$ of the previous subsection.) The other
parameters are fixed : $$ l_1 = (n-3)a_n l_2\,,\quad\quad
\ka =
\frac{1 -l_2 + (n-3)x_0^{'\,2}}{2x'_0}\,.$$ As $G(x)$ of (\ref{5-4n1c}) is
changed into
$G(x) - 2 a_n$ which is $< 0\,,$ now $\Om^2(x)$ monotonically decreases
from 1 to 0.
 Here again, up to an homothethy, there is one and only one
$S^1-S^{n-2}$ Einstein-Weyl regular metric, still with a positive scalar
conformal curvature. The metrics are the same, only the orientation of the
Einstein-Weyl manifold changes.

The particular case $l_2 = 1$ requires $x_0 = 0\,,\ l_1 = (n-3)a_n$ and
$\ka = (n-3)a_n/2\,.$

\subsubsection{Bolt($S^{n-2})$-Bolt($S^{n-2})$ metric}  Consider finally a
situation where the allowed range of
$x$ is 
$[x'_0,\,x_0]\,.$ From Lemma 8 and (\ref{n422b}), we know that a bolt(+1) 
at
$x'_0$ and a bolt(-1) at $x_0$ imply four relations between
$\ka,\ x_0,\ x'_0,\ l_1$ and $l_2\,:$ 
\beqa\label{7-4n1}
\ka  = -\frac{1 - l_2 +(n-3)x_0^{'\,2}}{2x'_0} & = & 
\frac{1 - l_2 +(n-3)x_0^{\,2}}{2 x_0} \,,\ l_2 \neq 1\nnb \\ 0 =
\Om^2(x'_0) & = & \Om^2(x_0)  \,.\eeqa  (The case  $l_2 = 1$ is  excluded
as the last two equations (\ref{7-4n1}) imply :
$ l_1 + (n-3)K_n(x'_0) = l_1 + (n-3)K_n(x_0) = 0$ which  enforces
$x_0 = x'_0 = l_1 = 0$ which is forbidden !) The first two equations lead
to two possibilities :
\beq\label{7-4n2} {\rm solution\ a)}\  :\ x_0 = -x'_0\ ;\quad\quad {\rm
solution\ b)}\ :\ x_0 x'_0 =
\frac{l_2-1}{n-3}\,.
\eeq Eliminating $l_1$ between the two others
 leads to the vanishing of a new function :
\beqa\label{7-4n3} h_n[x_0,x'_0] \equiv [\al_n(x_0) -
\al_n(x'_0)] & - & (n-3)\frac{l_2}{1-l_2}[K_{n}(x_0) - K_{n}(x'_0)] = 0\
\,, \\
\al_n(x) & = & \frac{1}{x(1+x^2)^{(n-4)/2}}\,.\nnb \eeqa
\nl\noindent The second square bracket in (\ref{7-4n3}) is positive. The
function
$\al_n$ is monotonically decreasing in the two
 domains $x<0$ and $x>0\,.$ So, if $x_0$ and $x'_0$ have the same sign,
$\dst\frac{l_2}{1-l_2}$ has to be negative ; on the other case, the first
square bracket in (\ref{7-4n3}) is positive and 
$\dst\frac{l_2}{1-l_2}$ has to be positive. Then, solution 
(\ref{7-4n2}-a)) requires $0 < l_2 < 1$ and solution (\ref{7-4n2}-b))
either $l_2 >1\,,$ the two zeroes of $\Om^2$ being of the same sign, or $0
< l_2 < 1$ when they are of opposite sign.

In both cases, $l_2
\le 0$ is excluded.

\vspace{3mm}
\begin{itemize}

\item {\it Solution b)}
\nl The derivative of the function $ h_n[x_0,x'_0(x_0)]$ is :
$$ \frac{dh_n[x_0,x'_0(x_0)]}{dx_0} = - \left(\frac{1}{x_0^2} +
\frac{n-3}{1-l_2}\right) \left[\frac{1}{(1+x_0^2)^{(n-2)/2}} -
\frac{1}{(1+x_0^{'\,2})^{(n-2)/2}}\right]\,.$$
\begin{itemize}
\item $l_2 >1$ : $x_0$ and $x'_0$ have the same sign (as ($
h_n[x_0,x'_0(x_0)]$ is an odd-parity function, we may chose a positive
sign ), then
$x_0 >  \sqrt{\dst\frac{l_2 -1}{n-3}}$ and $
\dst\frac{dh_n[x_0,x'_0(x_0)]}{dx_0}$ is negative : as a consequence,
$ h_n[x_0,x'_0(x_0)]$ decreasing from 0 when $x_0 =   
\sqrt{\dst\frac{l_2 -1}{n-3}}$ to $-\,\infty$ when $x_0$ goes to    
$+\,\infty$ does not vanish.
\item $0 < l_2 < 1$ : $x_0$ positive, and $ h_n[x_0,x'_0(x_0)]$ has a
minimum for $x_0 = - x'_0 =   \sqrt{\dst\frac{1-l_2}{n-3}}$ which is shown
to be positive when $l_2 \in ]\,0\,,\ 1\,[\,,$ that also excludes any
solution to (\ref{7-4n3}).
\end{itemize} Then we are left with case (\ref{7-4n2}-a)) 
\item {\it Solution a)} \nl $l_1 = 0$ results from the difference of the
last two equations (\ref{7-4n1}) with $x_0 = -x'_0\,.$ Moreover, with $0 <
l_2 < 1\,,$
$$\frac{dh_n[x_0,-x_0]}{dx_0} = -2\frac{1-l_2
+(n-3)x_0^2}{(1-l_2)x_0^2(1+x_0^2)^{(n-2)/2}}$$ is negative and
$ h_n[x_0,-x_0]\,,$ decreasing from $+\,\infty$ to
$-2l_2(n-3)a_n/(1-l_2)$  when
$x_0$ goes from $0$ to     
$+\,\infty\,,$ has a unique zero $x_0\,.$

\end{itemize} To sum up, given a positive parameter $l_2< 1\,,$ there is
one and only one bolt(+1)-bolt(-1) Einstein-Weyl regular metric with $l_1
=0$ and $\ka = \dst\frac{1-l_2 +(n-3)x_0^2}{2x_0}\,.$ 

 Note also that relation (\ref{7-4n2}-a) implies :
\beq\label{7-n22a}
 \om_3[x_0] = \om_3[x'_0] \,.\eeq

Moreover, as now :
\beq\label{5-4n1f}
\frac{d\Om^2}{dx} = (n-3)l_2[1+(n-3)x^2](1+x^2)^{(n-6)/2} [G(x) -
a_n]\,,\eeq decreases from $a_n$ to $-a_n\,,\ \Om^2(x)$ has a single
maximum between $-x_0$ and $x_0\,,$ precisely at $x=0$ as
$G(0) = a_n\,.$ As a consequence, 
\beq\label{s-3} 0 \le \Om^2(x) \le (1-l_2) < 1\ \Rightarrow
(n-3)l_2\frac{n\Ga}{2\ka}
\le S^D \le (l_2 +(n-4))\frac{n\Ga}{2\ka}\eeq  is positive on the
manifold, whatever the dimension $n \ge 4\,.$

\subsection{Summary}In our Gauduchon gauge, we found three, and only
three, families of non-conformally Einstein regular
 Einstein-Weyl metrics ;  according to the classification of Gibbons and
Hawking, they are complete and live on a compact manifold without boundary
; moreover, they have a  positive  conformal scalar curvature in agreement
with the
 theorem of Calderbank \cite{CalderbankFaraday}.

\section {Concluding remarks}  In this paper, we have first presented a
local analysis of n-dimensional  Einstein-Weyl structures ($g\,,\
\ga$) corresponding to cohomogeneity-one metrics in a Gauduchon  gauge. 
Second, we
have discussed with some details the explicit solutions in the case of an
$SU(m)$ group of left-isometries and in the case of an $S^1\times SO(n-1)$
group. 

In the first part, we
emphasized the role of the extra isometry exhibited by Tod, we explicited
its action for cohomogeneity-one structures, and we gave a necessary
condition for the existence of a non-exact Einstein-Weyl structure (Lemma
1) ; moreover, for a large subclass,we proved that the metric is locally
conformally K\"ahler (Theorem 2).

In the second part, we presented a complete local analysis of
the two families, we showed that they depend on 3 arbitrary
parameters (plus one homothetic one), we gave
the K\"ahler form (for a conformally related metric) for the first case
; then, in both cases, we analysed the consequences of the completeness
requirement and we obtained one-parameter families of solutions (plus one
homethetic parameter
$\Ga >0$).

\vspace{3mm}
 
Let us finally compare our results with previous ones. Of course, they are
not new when compared to global mathematical approaches, but here we
mainly required only local properties and so we obtained all the local
solutions. We also used a language more relevant for
physicists and, as in the search for special solutions we found a simpler
parametrisation, we were able to prove the conjectures in
\cite{Madsen-a} and to correct some mistakes in the 4-D analysis of
\cite{mpps}.

\begin{itemize}

\item As the analysis of Gibbons and Hawking
in the language of bolts and nuts applies only to orientable
manifolds, it is not surprising that we missed  metrics on non-orientable
manifolds such as
$RP^4$ or 
$RP^4 \# CP^2\,,$ contrarily to
\cite{{Madsen-a},{mpps}}.

\item There is a correspondance between Nuts and Bolts {\sl \`a la} 
Gibbons and Hawking \cite{GH} and special orbits in the language of
mathematicians :\begin{itemize}
\item a nut corresponds to special orbit being a point,
\item a bolt(p) in n =  2m dimensions, corresponds to special orbit being
$\mathbb{C}P^{m-1}\,;$ 

the integer $p\quad ( p< m )$ means that the original
(n-1)-dimensional homogeneous space $SU(m)/SU(m-1)$ has been
restricted, through Einstein-Weyl constraints and regularity
requirements, to $((U(1)/\mathbb{Z}_p)\times SU(m))/U(m-1)\,,$ 
\item a bolt($S^1)$ corresponds to special orbit being a circle,
\item a bolt($S^{n-2})$ corresponds to special orbit being a 
(n-2)-dimensional sphere.
\end{itemize}

\item Our nut-bolt families  (Subsects. 4.2.2 and  4.2.3) correspond to the
same manifold $\mathbb{C}P^{m}$ with both orientations : so the solutions
are not really different solutions. The same remark also holds for the
bolt($S^1)$-bolt($S^{n-2})$ solutions of Subsects. 5.2.2 and 5.2.3, the
manifold being $S^n\,.$

\item Our bolt-bolt families (Subsects. 4.2.4 and 5.2.4) corresponds to
Madsen's ones \cite[Subsects. 8-24 and 7-40]{Madsen-a}, but we have been
able to prove that for structures of cohomogeneity-one under $SU(m)\,,$
 no bolt(p)-bolt(p) exists with
$p\ge m$ (Lemma 5), a result which was only conjectured. 
Moreover, thanks to our parametrisation that disentangles the parameters
$\ka,\ x_0\ $ and $x'_0$ into a single transcendental equation for only one
unknown parameter, we also proved that the relation conjectured in
Madsen's thesis dissertation ( the ``time parameter" in these analyses
being an angle $\varphi) \ : \Phi_1 + \Phi_2 = \pi \Leftrightarrow h^2(T_0)
= h^2(T^{\,'}_0)
\Leftrightarrow \om_3(x_0) = \om_3(x^{\,'}_0) $ (c.f.
(\ref{7-n2a},\ref{7-n22a})) , is indeed the
sole solution, for any $n \ge 4\,.$

\end{itemize}

{\bf \large Aknowledgements :} it is a pleasure to thank Fran\c{c}ois
Deduc for helpful discussions and the referee for pointing some uncorrect assertions in the manuscript. 

\vspace{3cm}

\section {Appendix A : Cohomogeneity-one geometry}  The n-dimensional
distance being split into 
$$ds^2 = (dT)^2 + h_{ij}[T] e^i e^j = (dT)^2 +
g_{\al\be}dx^{\al}dx^{\be}$$ and using the quantities $K_i^j$ given in
(\ref{2-12}), the Christoffel connection components are expressed as :
\beqa\label{A-0} 2\Ga^{\al}_{0\be} = E^{\al}_ie^j_{\be} K_j^i\ \ & , &
\ \ 2\Ga^{0}_{\al\be} = - e_{\al}^ie^j_{\be} K_{ij}\ \ ,\nnb \\
2\Ga^{\al}_{\be\ga} =  g^{\al\de}[g_{\de\be\,,\ga} + g_{\de\ga\,,\be} -
g_{\be\ga\,,\de}]\ \ & , & \ \ {\rm the\ other\ components\ vanishing} 
\\ {\rm with}\ \ g^{\al\be}g_{\be\ga} = \de^{\al}_{\ga}\ & , &
E_i^{\al}e^j_{\al} = \de_i^j\ \,,\ \ E_i^{\al}e^i_{\be} =
\de^{\al}_{\be}\ .\nnb  
\eeqa \noindent The covariant derivative of the vielbeins
$e^i_{\al}$ is readily computed :
\beqa\label{A-1}
\nabla_{\be}e^i_{\al} & = &  \partial_{\be}e^i_{\al} -
\Ga^{\ga}_{\be\al}e^i_{\ga}\nnb \\ & = &\partial_{\be}e^i_{\al} -
\frac{1}{2}h^{il}h_{mn}E^{\ga}_l[\partial_{\be}(e^m_{\al}e^n_{\ga}) +
\partial_{\al}(e^m_{\be}e^n_{\ga}) -
\partial_{\ga}(e^m_{\al}e^n_{\be}) ]\,, \eeqa  which, using
(\ref{2-110a},\ref{014}), reduces to
\beq\label{A-2} \nabla_{\be}e^i_{\al} = -
\left[\frac{1}{2}f^i_{jk}e^j_{\be} + f^i_{ak}\om^a_{\be}\right]e^k_{\al} +
h^{ij}h_{k(l}f^k_{n)j}e^l_{\al}e^n_{\be}\,.
\eeq

A related result is :\beq\label{016}
\nabla_{\al}E_i^{\al} = f_{ki}^k + \om^a_{\be}E^{\be}_k\,f^k_{ai}\,,\eeq

\noindent With  (\ref{A-0}), the n-dimensionnal Ricci tensor is expressed
in function of the tensor $h_{ij}[T]$, its derivative
$K_{ij}$ and the (n-1)-dimensional Ricci tensor (\ref{2-12}). The 
expression
$$2 R_{0\al}^{(\nabla)} = \nabla_{\be}(e^i_{\al}E^{\be}_jK_i^j) -
\nabla_{\al}(e^i_{\be}E^{\be}_jK_i^j) $$ simplifies to
$$K_i^j[T]\nabla_{\be}(e^i_{\al}E^{\be}_j)$$ which, using
(\ref{2-110a},\ref{A-2}), reduces to :
\beq\label{A-3}
 2R_{0\al}^{(\nabla)} = e^i_{\al}[K_{i}^jf_{kj}^k + K_{k}^jf_{ij}^k]
\,.\eeq

\subsection{The Bianchi identity} The $\nu = 0$ component of the Bianchi
identity
$2\nabla_{\mu}R^{(\nabla)\,\mu}_{\nu} = \nabla_{\nu}R^{(\nabla)}$ is split
according to $\mu = (0\,,\al)\ .$ Using (\ref{2-12},\ref{A-1}) and
$$R^{(\nabla)} = R^{(n-1)} + 2R^{(\nabla)}_{00}
-\frac{1}{4}(\frac{h'}{h})^2 +
\frac{1}{4}K_{ij}K^{ij}$$ one obtains :
$$2\nabla_{\mu}R^{(\nabla)\,\mu}_{0} = \nabla_{0}R^{(\nabla)}
-h^{ij}\frac{dR^{(n-1)}_{ij}}{dT} +
2\nabla_{\al}^{(n-1)}R^{(\nabla)\,\al}_{0}\,.$$ As a consequence :
\beq\label{A-4} h^{ij}\frac{dR^{(n-1)}_{ij}}{dT} = 2[\nabla_{\al}
E^{\al}_k]R^{i}_{0}\,,
\eeq where, with (\ref{A-3}), $$2R^{i}_{0} =
2h^{ij}E_j^{\al}R^{(\nabla)}_{0\al} = K^{ji}f_{kj}^k +
K_k^jh^{il}f_{lj}^k\,.$$

\section {Appendix B : Bolt-Nut versus Nut-Bolt}

The relation (\ref{7-n2} -a)) whose particular case is (\ref{6-n3})
implies :
$$ \frac{x_0 + x'_0}{1 -x_0 x'_0} = \frac{-1}{\ka}$$ ($\ka$ being related
to $x_0$ and $x'_0$ through (\ref{7-n2})). With
$$\phi_0 = \tan^{-1}(x_0) \ ,\ \ \phi'_0 = \tan^{-1}(x'_0) \ ,\ \
\psi =
\tan^{-1}(\ka) \ ,\ \ \phi_0, \phi'_0 \in ]-\pi/2,\,+\pi/2[\,,\
\psi \in ]\phi_0,\,+\pi/2[\,,$$ this relation writes
\beq\label{a-1}
\psi - \phi_0 = \pi/2  + \phi'_0\,.\eeq The following identity :
\beq\label{a-2} H_n(\ka,\,x_0) \equiv
\int_{-\,\infty}^{x_0}\frac{[\ka - y]^{n-1}}{(1+ y^2)^{n}}dy =
\int_{x'_0}^{\ka}\frac{[\ka - z]^{n-1}}{(1+ z^2)^n}dz \equiv
\tilde{H}_n(\ka,\,x'_0)
\eeq is proven after the change of integration variables :$y =
\tan(\Phi)\,,\ z = \tan(\psi -\pi/2 - \Phi)\,.$ Note that the same
manipulations give no information on the similar integral between
$x_0$ and $x'_0\,.$

The function $g_m[\ka, x_0]$ of (\ref{5-n4}) may be expressed as :
\beqa\label{a-3} g_m[\ka, x_0]&  = & 2m\frac{\ka +
l_2}{1+\ka^2}H_{m+1}(\ka,\,x_0) -[m-1
+\frac{m\ka(\ka+l_2)}{1+\ka^2}]H_m(\ka,\,x_0) +\nnb\\ & + &
\frac{(\ka-x_0)^{m-2}}{2m(1+x_0^2)^{m-1}}\left[m-1 -\frac{2m(\ka+l_2)(\ka
-x_0)(1+\ka x_0)}{(1+\ka^2)(1+x_0^2)}\right]\,.
\eeqa In the same manner, the function $g'_m[\ka, x'_0]$ of (\ref{6-n2})
may be expressed as :
\beqa\label{a-4} g'_m[\ka, x'_0]&  = & 2m\frac{\ka +
l_2}{1+\ka^2}\tilde{H}_{m+1}(\ka,\,x'_0) -[m-1
+\frac{m\ka(\ka+l_2)}{1+\ka^2}]\tilde{H}_m(\ka,\,x'_0) -\nnb\\ & - &
\frac{(\ka-x'_0)^{m-2}}{2m(1+x_0^{'\,2})^{m-1}}\left[m+1
-\frac{2m(\ka+l_2)(\ka -x'_0)(1+\ka
x'_0)}{(1+\ka^2)(1+x_0^{'\,2})}\right]\,.
\eeqa So, using the identity $$\frac{\ka -x_0}{1+x_0^2} =
\frac{\ka -x'_0}{1+x_0^{'\,2}}$$ resulting from (\ref{6-n3}), one gets
$$g_m[\ka, x_0] + g'_m[\ka, x'_0]
 = 0\,. \qq\qq  {\rm Q.\ E.\ D.}$$

\vspace{3cm} 
\bibliographystyle{plain}

\begin {thebibliography}{49}

\bibitem{OR} L. O'Raifeartaigh, {\sl ``The dawning of gauge theory"},
Princeton University Press, Princeton, N.J. (1997).

\bibitem{HW29} H. Weyl, {\sl `` Gravitation und Elektrizit\"at"}, {\sl Sitzungsber. Preuss. Akad. Berlin} (1918) 465
; {\sl ``Space-Time-Matter"}, Methuen London (1922) ; {\sl ``Elektron und  Gravitation"},  {\sl Zeit. f.
Physik} {\bf 330} (1929) 56 and the book
\cite{OR}.

\bibitem{JTH} J. T. Wheeler, {\sl ``New conformal gauging and the electromagnetic theory of Weyl"}, {\sl J. Math. Phys.} {\bf 39} (1998)  299.

\bibitem{WD}  W. Drechsler and H. Tann,  {\sl ``Broken Weyl invariance and the origin of mass}", gr-qc/9802044, {\sl Found. Phys.}
{\bf 29} No.7 (1999) 1023 ;  W. Drechsler,  {\sl ``Mass-generation by Weyl-symmetry  breaking}", gr-qc/9901030, {\sl Found.
Phys.} {\bf 29} No.9 (1999) in press.
 
\bibitem{MHE}  K. Meetz, {\sl  ``Realization of chiral symmetry in a curved isospin space}", {\sl J. Math. Phys.} {\bf 10} (1969) 65 ; J.
Honerkamp, {\sl  ``Chiral multi-loops"}, {\sl Nucl. Phys.} {\bf B36} (1972) 130 ; G. Ecker and J.
Honerkamp,  {\sl `` Application of the invariant renormalisation to the non-linear chiral invariant pion lagrangian in the one-loop approximation"},  {\sl Nucl. Phys.} {\bf B35} (1971) 481.

\bibitem{Calderbank} D.M.J. Calderbank and H. Pedersen, {\sl
``Einstein-Weyl geometry"}, Edinburgh Preprint MS-98-010,  "Essays on
Einstein manifolds", Cambridge, MA, International Press.

\bibitem{papado-valent} G. Papadopoulos,  {\sl ``Elliptic monopoles and (4,0) supersymmetric sigma models with torsion"}, {\sl Phys. Lett.} {\bf 356B}
(1995)  249 ;\nl T. Chave, K. P. Tod and G. Valent, {\sl ``(4,0) and (4,4) $\si $ models with a triholomorphic Killing vector"}, {\sl Phys. Lett.} {\bf 383B} (1996)  262. 

\bibitem{Tod96} K. P. Tod, {\sl ``Local heterotic geometry and self-dual Einstein-Weyl spaces"}, {\sl Class. Quantum Grav.} {\bf 13} (1996) 2609.

\bibitem{PS93} H. Pedersen and A. Swann, {\sl ``Riemannian submersions, four-manifolds and Einstein-Weyl geometry"}, {\sl Proc. Lond. Math. Soc.} {\bf
66} (1993) 381.

\bibitem{HullWitten85} C. M. Hull and E. Witten, {\sl ``Supersymmetric sigma models and the heterotic string"}, {\sl Phys. Lett.} {\bf
160B} (1985)  398 ; C. M. Hull, {\sl `` Sigma model beta-functions and  string compactifications"}, {\sl Nucl. Phys.} {\bf B267} (1986) 266.

\bibitem{xxy} E. Bergshoeff and E. Sezgin, {\sl ``The (4,0) heterotic string with Wess-Zumino term"}, {\sl Mod. Phys. Lett.} {\bf A1}
(1986) 191 ;
\newline P. Howe and G. Papadopoulos, {\sl ``Ultraviolet behavior of two-dimensional supersymmetric nonlinear sigma models"},
{\sl Nucl. Phys.} {\bf B289} (1987)
264 ; {\sl ``Anomalies in two-dimensional supersymmetric nonlinear sigma m odels"}, {\sl Class. Quantum Grav.} {\bf 4} (1987) 1749 ; {\sl ``Further remarks on the geometry of two-dimensional nonlinear sigma models"}, {\sl Class. Quantum
Grav.} {\bf 5} (1988) 1647 ; \newline Ph. Spindel, A. Sevrin, W. Troost
and A. Van Proyen, {\sl ``Extended 
supersymmetric sigma models on group manifolds. 1. The complex structures"}, {\sl Nucl. Phys.} {\bf B308} (1988) 662.

\bibitem{Delduc} F. Delduc and G. Valent, {\sl ``New geometry from heterotic supersymmetry "}, {\sl Class. Quantum Grav.} {\bf
10} (1993) 1201.


\bibitem{Bonneau97} G. Bonneau, {\sl ``Einstein-Weyl structures  and Bianchi metrics"}, {\sl Class. Quantum Grav.} {\bf 15} 
(1998) 2415.

\bibitem{Bonneau98} G. Bonneau,  {\sl ``Compact Einstein-Weyl four-dimensional manifolds"}, {\sl Class. Quantum Grav.} {\bf 16} 
(1999) 1057.

\bibitem{Gauduchon} R. Gauduchon, {\sl ``La 1-forme de torsion d'une vari\'et\'e hermitienne compacte"}, {\sl Math. Ann.} {\bf 267} (1984) 495.


\bibitem{Gauduchonbis} P. Gauduchon, {\sl ``Structures de Weyl-Einstein, espaces de twisteurs et vari\'et\'es de type $S^1 \times S^3$ "},  {\sl J. Reine Angew. Math. } {\bf
469} (1995) 1.

\bibitem{Tod92} K. P. Tod, {\sl `` Compact 3-dimensional Einstein-Weyl structures"}, {\sl J. London Math.Soc. 2}{\bf 45}(1992) 341.

\bibitem{PedTod92} H. Pedersen and K. P. Tod, {\sl ``Three-dimensional Einstein-Weyl geometry"}, {\sl Adv. Math.} {\bf
97}(1993) 74.



\bibitem{PDSRheine} H. Pedersen and A. Swann, {\sl ``Einstein-Weyl geometry, the Bach tensor and conformal scalar curvature"}, {\sl J. Reine Angew. Math. }
{\bf 441} (1993) 99.

\bibitem{Madsen-a} A. Madsen, {\sl ``Compact Einstein-Weyl manifolds with
large symmetry group"}, PhD. Thesis, Odense University, 1995.

\bibitem{mpps} A.B. Madsen, H. Pedersen, Y.S. Poon and A.F. Swann, {\sl ``Compact Einstein-Weyl manifolds with large symmetry group"},  {\sl
Duke Math. J.} {\bf 88} (1997) 407.

\bibitem{CalderbankFaraday}  D.M.J. Calderbank,  {\sl ``The Faraday 2-form
in Einstein-Weyl geometry"}, to appear in {\sl  Math. Scand. } (2000).
 
\bibitem{GH} G. W. Gibbons and S. W. Hawking, {\sl ``Classification of gravitational instanton symmetries"}, {\sl Commun. Math. Phys.
}{\bf 66} (1979) 291 ; G. W. Gibbons and C. N. Pope,  {\sl ``The positive action conjecture and asymptotically euclidean metrics in  quantum gravity"}, {\sl Commun. Math.
Phys. }{\bf 66} (1979) 267.


\bibitem{Besse} A. L. Besse, {\sl ``Einstein Manifolds"}, Springer Verlag
1987, [7.12, p.179].

\bibitem{Friedan} D. Friedan, {\sl ``Non-linear models in $2+\epsilon$  dimensions"}, {\sl Annals of Phys. (NY)}{\bf 163} (1985) 
318. 


\bibitem{KoNo} S. Koyabashi and K. Nomizu, {\sl ``Foundations of
Differential Geometry"}, Interscience Publishers, John Wiley, 1969 [Vol.
II, p. 199]. 
\bibitem{BBBCD} C. Becchi, A. Blasi, G. Bonneau, R. Collina and F. Delduc,
{\sl ``Renormalizability and infrared finiteness of non-linear $\si$  models~: a regularization-independant analysis for compact coset  spaces"}, {\sl Comm. Math. Phys.} {\bf 120} (1988) 121.

\bibitem{Landau} L. D. Landau and E. M. Lifshitz, {\sl ``The Classical
Theory of Fields"}, 4th edition (Oxford : Pergamon Press, 1975).

\bibitem{russes} V. A. Belinskii et al., {\sl ``A general solution of the Einstein equations with a time singularity"}, {\sl Advances in Phys.} {\bf 31}
(1982) 639.

\bibitem{W} S. T. Weinberg, {\sl ``Gravitation and Cosmology :  principles and applications to the general theory of relativity "},
Willey, New York (1972), chapter
13.

\bibitem{Bonneau} G. Bonneau, {\sl ``Einstein-Weyl structures coresponding to diagonal K\"ahler Bianchi IX metrics"}, {\sl Class. Quantum Grav.} {\bf 14} (1997)
2123.

\bibitem{Kahlersymmetrique} A. Borel, {\sl ``K\"ahlerian coset spaces of semisimple Lie groups"}, {\sl Proc. Natl. Acad. Sci.} {\bf
40} (1954) 1147 ;\nl A. Borel and F. Hirzebruch, {\sl ``Characteristic classes and homogeneous spaces"}, {\sl American J. Math.}
{\bf 80} (1958) 458. 

\bibitem{Vaisman} I. Vaisman, {\sl ``Generalised Hopf manifolds"}, {\sl Geom. Dedicata} {\bf 13} (1982) 231.

\bibitem{Madsen-b} A. Madsen, {\sl ``Einstein-Weyl structures in the conformal classes of LeBrun metrics"}, {\sl Class. Quantum Grav.} {\bf 14} (1997)
2635.


\bibitem{DS95} A. S. Dancer and Ian A. B. Straham, {\sl Cohomogeneity-One
K\"ahler metrics} in ``Twistor theory", S. Huggett ed., Marcel Dekker
Inc., New York, 1995, p.9.

\bibitem{Tod95} K. P. Tod, {\sl  Cohomogeneity-One metrics with Self-Dual
Weyl tensor} in ``Twistor theory", S. Huggett ed., Marcel Dekker Inc., New
York, 1995, p.171.


\end {thebibliography}

\end{document}